\documentclass[%
 reprint,
superscriptaddress,
nofootinbib,
 aps,
 twocolumn,
 prd
]{revtex4-2}

\usepackage{bm}
\usepackage[english]{babel}
\usepackage{amsmath}
\usepackage{amsthm}
\usepackage{amssymb}
\usepackage[colorlinks = true,breaklinks=true]{hyperref}
\usepackage{tensor}

\usepackage{graphicx}
\usepackage[capitalise]{cleveref}
\usepackage{acro}
\usepackage[normalem]{ulem}

\newcommand{\hnabla}{\overset{\mathrm{h}}{\nabla}}
\newcommand{\vnabla}{\overset{\mathrm{v}}{\nabla}}

\numberwithin{equation}{section}
\renewcommand\theequation{\arabic{section}.\arabic{equation}}

\begin{document}

\title{Semiclassical analysis of Dirac fields on curved spacetime}

\author{Marius A. Oancea}
\email{marius.oancea@univie.ac.at}
\address{Faculty of Physics, University of Vienna, Boltzmanngasse 5, 1090 Vienna, Austria}

\author{Achal Kumar}
\email{achalkumar@iisc.ac.in}
\address{Department of Physics, Indian Institute of Science, Bangalore 560012, India}

\begin{abstract} 
We present a semiclassical analysis for Dirac fields on an arbitrary spacetime background and in the presence of a fixed electromagnetic field. Our approach is based on a Wentzel-Kramers-Brillouin approximation, and the results are analyzed at leading and next-to-leading order in the small expansion parameter $\hbar$. Taking into account the spin-orbit coupling between the internal and external degrees of freedom of wave packets, we derive effective ray equations with spin-dependent terms. These equations describe the gravitational spin Hall effect of localized Dirac wave packets. We treat both massive and massless Dirac fields and show how a covariantly defined Berry connection and the associated Berry curvature govern the semiclassical dynamics. The gravitational spin Hall equations are shown to be particular cases of the Mathisson-Papapetrou equations for spinning objects.
\end{abstract}

\maketitle

\section{Introduction}

Semiclassical dynamics represents an intermediate regime between classical and quantum mechanics, where wave effects such as diffraction and interference can be ignored and the average dynamics of wave packets can be well approximated by point particles \citep{Berry_1972,Herman1994}. While wave dynamics is generally described by partial differential equations, semiclassical approximations can be used to obtain an effective description in terms of point particles with dynamics governed by ordinary differential equations. 

It is well known from optics and condensed matter physics that wave packets with internal degrees of freedom, such as polarization or spin, can have nontrivial dynamics in the semiclassical limit. In particular, spin-orbit interactions between the external (average position and momentum) and internal (spin or polarization) degrees of freedom of the wave packet can lead to spin Hall effects \citep{SHE_review, SHE_review1, SOI_review}. In this case, the propagation of the wave packets becomes spin dependent. These effects have been studied theoretically and observed in many experiments for the propagation of polarized light beams in optical media \citep{SHE-L_original,SHE_original,Onoda2006,Duval2006,Duval2007,SHEL_experiment,Bliokh2008,Bliokh2009,SHEL_experiment3,SHEL_experiment4}, as well as for electrons in condensed matter systems \citep{originalSHE1,originalSHE2,originalSHE3,originalSHE4}.

Spin Hall effects are also expected to occur for wave packets propagating in gravitational fields, an effect known as the gravitational spin Hall effect \citep{GSHE_review,Oanceathesis}. Electromagnetic \citep{GSHE2020,Frolov2020,Harte_2022,SHE_QM1} and linearized gravitational \citep{GSHE_GW,gshe_lensing,SHE_GW} wave packets have been shown to follow frequency- and polarization-dependent trajectories on curved spacetimes. Similar effects have also been shown for massive Dirac wave packets propagating on curved spacetimes \citep{audretsch,audretsch_torsion,rudiger,SHE_Dirac,Cianfrani_2008,CIANFRANI2008} (see also Refs. \citep{Palmer2012,Laszlo2018,obukhov2001,obukhov2009,obukhov2011,obukhov2013spin,obukhov2017general}). Furthermore, spin-dependent effects in curved spacetime also play an important role in chiral kinetic theory \citep{liu2018chiral,PhysRevD.105.096019,KAMADA2022104016,Liu_2020}, which is a semiclassical approach to studying many-body effects arising from interactions in dilute gases of microscopic chiral fermions. Other, more exotic, spin Hall effects have also been recently studied for massless particles with anyonic spin \citep{marsot2022,gray2022,marsot2022hall}. 

In this paper, we study the semiclassical dynamics of charged Dirac fields in curved spacetime and in the presence of a fixed electromagnetic field. We consider both massive and massless fields and treat them separately since their behavior is radically different. Our approach is based on a Wentzel-Kramers-Brillouin (WKB) approximation, where the reduced Planck constant $\hbar$ is taken as the small expansion parameter. By taking into account the spin-orbit interactions between the external and internal degrees of freedom of the wave packets, we derive spin-dependent equations of motion describing the gravitational spin Hall effect. The semiclassical dynamics we derive is based on a covariantly defined Berry connection (which determines the dynamics of the spin internal degree of freedom) and the associated Berry curvature (which determines the dynamics of the average external degrees of freedom). Furthermore, we show that the gravitational spin Hall equations for both massive and massless Dirac wave packets can be viewed as particular cases of the Mathisson-Papapetrou equations for spinning objects. We need to mention that similar results, based on WKB approximations, have also been obtained in Refs. \citep{audretsch,rudiger,SHE_Dirac,Cianfrani_2008,CIANFRANI2008}. However, these studies are limited to massive Dirac fields, and no external electromagnetic field is considered.

The paper is organized as follows. We start in Sec. \ref{sec:Dirac_eq} by introducing Dirac fields, the Dirac equation on curved spacetime and a variational formulation of the Dirac equation in terms of an action. In Sec. \ref{sec:wkb}, we define our WKB approximation. The semiclassical expansion in powers of $\hbar$ is performed at the level of the action, and then the semiclassical WKB equations are derived as the Euler-Lagrange equations. In Sec. \ref{sec:massive}, we consider the case of massive Dirac fields and analyze the semiclassical WKB equations. We study the algebraic properties of the principal symbol of the Dirac operator, we introduce the Berry connection and the Berry curvature, and then we take into account spin-orbit interactions in order to derive the gravitational spin Hall equations. These are shown to be a particular case of the Mathisson-Papapetrou equations. The same analysis is performed in Sec. \ref{sec:massless} for massless Dirac fields. In addition, here we also give a comparison with similar known results in the context of the gravitational spin Hall effects for electromagnetic and gravitational waves. Finally, we present our conclusions in Sec. \ref{sec:conclusions}.

\textit{Notation and conventions:} We consider an arbitrary smooth Lorentzian manifold $(M, g_{\mu \nu})$, where the metric tensor $g_{\mu \nu}$ has the signature $(-\,+\,+\,+)$. The absolute value of the metric determinant is denoted as $g = |\det g_{\mu \nu}|$. Phase space is defined as the cotangent bundle $T^*M$, with canonical coordinates $(x, p)$.  Furthermore, on $M$ we consider a fixed electromagnetic field $F_{\mu \nu} = 2 \nabla_{[\mu} A_{\nu]}$. The Einstein summation convention is assumed, and we use the notation $a_\alpha b^\alpha = a \cdot b$. Greek indices, $(\alpha, \beta, ...)$, represent spacetime indices, and run from $0$ to $3$. Latin indices, $(a, b, ...)$, represent tetrad indices and run from $0$ to $3$. The components of $3$-vectors are denoted using Latin indices from the middle of the alphabet, $(i, j, ...)$, which run from $1$ to $3$. Eigenspinors will be labeled with capital Latin indices, $(A, B, ...)$, which run from $0$ to $1$. We also use a summation convention for repeated eigenspinor indices.

\section{The Dirac equation} \label{sec:Dirac_eq}

Consider a Lorentzian manifold $(M, g_{\mu \nu})$, which is a solution of the Einstein field equations, admitting a spin structure \citep[p. 416]{MR685274}. A Dirac field $\Psi$ is a section of a vector bundle with fiber $\mathbb{C}^4$, associated with the spin frame principal bundle $\mathrm{Spin}_{3,1}(M)$ via the representation $\rho(\Lambda) = \Lambda$, where $\Lambda \in \mathrm{Spin}(3, 1) = \mathrm{SL}(2, \mathbb{C})$ \citep[p. 418]{MR685274}. The Dirac field $\Psi$, of charge $e$ and mass $m$, satisfies the Dirac equation
\begin{equation} \label{eq:Dirac_eq}
    \left( i \hbar \gamma^\mu \nabla_\mu - e \gamma^\mu A_\mu  - m \right) \Psi = 0,
\end{equation}
where $A_\mu$ is the electromagnetic vector potential and $\gamma^\mu$ are the spacetime gamma matrices. These are related to the flat spacetime gamma matrices $\gamma^a$ by the tetrad fields $(e_a)^\mu$: $\gamma^\mu = (e_a)^\mu \gamma^a$. The spinor covariant derivative $\nabla_\mu$ is defined by a spin connection on the spin frame bundle $\mathrm{Spin}_{3,1}(M)$. Given a spin structure on $M$, the Levi-Civita connection on the Lorentz frame bundle $\mathrm{L}(M)$ determines a spin connection on the spin frame bundle $\mathrm{Spin}_{3,1}(M)$ \citep[p. 419]{MR685274}. The spinor covariant derivative $\nabla_\mu$ acts on spinor fields as
\begin{equation} \label{eq:cov_der}
    \nabla_\mu \Psi = \left( \partial_\mu - \frac{i}{4} \omega\indices{_\mu^{a b}} \sigma_{a b} \right) \Psi,
\end{equation}
where $\sigma_{a b} = \frac{i}{2} [\gamma_a, \gamma_b]$, and $\omega\indices{_\mu^{a b}}$ is the spin connection, defined as
\begin{equation}
    \omega\indices{_\mu^{a b}} = (e^a)_\nu \nabla_\mu (e^b)\indices{^\nu}. 
\end{equation}
The Dirac equation can also be derived from the following action:
\begin{equation} \label{eq:Dirac_action}
J = \int_{M} \mathrm{d}^4 x \, \sqrt{g}\, \Bar{\Psi} \hat{D} \Psi ,
\end{equation}
where $\Bar{\Psi} = \Psi^\dagger \gamma^0$, and the Dirac operator is
\begin{equation}
    \hat{D} = i \hbar \gamma^\mu \nabla_\mu - e A_\mu \gamma^\mu - m.
\end{equation}
Since the action is invariant under $U(1)$ transformations $\Psi \mapsto e^{i \theta} \Psi$, the following Dirac vector current $j^\mu$ is conserved:
\begin{equation} \label{eq:Dirac_current}
     j^\mu = \bar{\Psi} \gamma^\mu \Psi, \qquad \qquad \nabla_\mu j^\mu = 0.
\end{equation}
In the massless case ($m = 0$), the action admits an additional symmetry given by the transformation
\begin{equation}
    \Psi \mapsto e^{i \phi \gamma^5} \Psi, \qquad \bar{\Psi} \mapsto \bar{\Psi} e^{i \phi \gamma^5}.
\end{equation}
This symmetry gives the following conserved axial vector current:
\begin{equation}
    j_{axial}^\mu = \bar{\Psi} \gamma^\mu \gamma^5 \Psi, \qquad \qquad \nabla_\mu j_{axial}^\mu = 0.
\end{equation}

\section{WKB approximation} \label{sec:wkb}

We assume that the Dirac field admits a WKB expansion of the form
\begin{equation} \label{eq:WKB_Dirac}
\begin{split}
    \Psi (x) &=  \psi(x, \nabla_\mu S, \hbar) e^{ i S(x) / \hbar}, \\
    \psi(x,  \nabla_\mu S, \hbar) &= {\psi_0}(x,  \nabla_\mu S) + \hbar {\psi_1}(x,  \nabla_\mu S) + \mathcal{O}(\hbar^2),
\end{split}
\end{equation}
where $S$ is a real scalar function, $\psi$ is a complex amplitude spinor, and Planck's constant $\hbar$ represents a small expansion parameter. Note that we are allowing the amplitude $\psi$ to depend on $\nabla_\mu S$. This is justified by the mathematical formulation of the WKB approximation \citep{MR1806388,Emmrich1996}, where $\nabla_\mu S$ determines a Lagrangian submanifold $x \mapsto (x, \nabla_\mu S(x)) \in T^*M$, and the amplitude $\psi$ is defined on this Lagrangian submanifold. 

The main assumption behind this approximation is that the length scales of variation of the background spacetime and background electromagnetic field are much larger than the typical length scale of our wave packets, taken to be proportional to $\hbar$. 

The semiclassical analysis of the Dirac equation is usually performed by inserting the WKB ansatz \eqref{eq:WKB_Dirac} into the Dirac equation \eqref{eq:Dirac_eq} and analyzing the results order-by-order in $\hbar$ \citep{Rubinow-Keller,audretsch,rudiger,Stone2015(2)}. However, we find it more convenient to perform the semiclassical analysis at the level of the action \eqref{eq:Dirac_action}. The advantages of this variational approach are extensively discussed in Ref.~\citep{tracy2014}.

A variational formulation of the WKB approximation for the Dirac field is obtained by inserting the WKB ansatz \eqref{eq:WKB_Dirac} into the action \eqref{eq:Dirac_action}. We obtain
\begin{equation} \label{eq:action_WKB1}
\begin{split}
    J &= \int_M  \mathrm{d}^4 x \, \sqrt{g} \, \left( \bar{\psi} e^{-i S/ \hbar} \right) \hat{D} \left( \psi e^{i S/ \hbar} \right) \\
    &= \int_M  \mathrm{d}^4 x \, \sqrt{g} \, \bar{\psi} \left( D + i \hbar \gamma^\mu \nabla_\mu \right) \psi + \mathcal{O}(\hbar^2) .
\end{split}
\end{equation}
where
\begin{subequations}
\begin{align}
    D &= -\gamma^\mu k_\mu - m \mathbb{I}_4, \\
    k_\mu &= \nabla_\mu S + e A_\mu.
\end{align}
\end{subequations}
The action depends on the phase function $S(x)$ and the amplitudes $\psi(x,\nabla S)$ and $\bar{\psi}(x,\nabla S)$. Performing a variation of the action with respect to these fields (since the amplitude $\psi$ depends on $\nabla S$, the variation of the action must be performed as in \citep[Appendix B]{GSHE2020}), we obtain the following Euler-Lagrange equations:
\begin{subequations} \label{eq:HJ_full}
\begin{align}
    D \psi + i \hbar \gamma^\mu \nabla_\mu \psi &= \mathcal{O}(\hbar^2), \label{eq:HJ1_fullA} \\
    \bar{\psi} D - i \hbar (\nabla_\mu \bar{\psi}) \gamma^\mu &= \mathcal{O}(\hbar^2), \label{eq:HJ2_fullA} \\
    \nabla_\mu \left( \bar{\psi} \gamma^\mu \psi \right) &= \mathcal{O}(\hbar^2) \label{eq:transp_fullA}.
\end{align}
\end{subequations}
Equations \eqref{eq:HJ1_fullA} and \eqref{eq:HJ2_fullA} can also be obtained by directly inserting the WKB ansatz into the Dirac equation, and Eq.~\eqref{eq:transp_fullA} represents the WKB approximation of the conservation law given in Eq. \eqref{eq:Dirac_current}. 

We continue by analyzing the above equations at each order in $\hbar$. However, we have to treat the massive and massless cases separately. The main properties of semiclassical dynamics are governed by the null eigenspace of the principal symbol matrix $D$ \citep{Emmrich1996}, and there will be significant differences between the massive and massless case.

\section{Massive Dirac fields} \label{sec:massive}

In this section, we present our semiclassical analysis of massive Dirac fields. Our main goal is to understand the behavior of the massive Dirac field beyond the leading-order approximation and to obtain the equations of motion describing the gravitational spin Hall effect. We start in Sec. \ref{sec:0thGO} by examining the WKB equations at the lowest order in $\hbar$. We obtain the lowest-order dispersion relation, a transport equation for the intensity of the field, and we analyze the null eigenspace of the principal symbol matrix $D$. In Sec. \ref{sec:ray0}, we treat the dispersion relation as a Hamilton-Jacobi equation and solve it using the method of characteristics. We obtain that, at the leading order in the WKB expansion, massive Dirac fields can be described by charged massive point particles satisfying the covariant Lorentz force law. In Sec. \ref{sec:WKB_1_massive}, we analyze the WKB equations at order $\hbar^1$. We obtain a transport equation for the amplitude $\psi_0$, and we show how this can be expressed in terms of a Berry connection. In Sec. \ref{sec:geometry_Berry}, we present a geometric discussion, showing how the Berry connection can be viewed as a $\mathfrak{u}(2)$-valued one-form on the Lagrangian submanifold, and we calculate the associated Berry curvature. In Sec. \ref{sec:eff_disp}, we derive an effective dispersion relation, taking into account spin-orbit couplings, and we use this to derive spin-dependent ray equations, representing the gravitational spin Hall effect. Finally, in Sec. \ref{sec:MPD_comp}, we compare these ray equations with the Mathisson-Papapetrou equations and with other known results from the literature.

\subsection{WKB equations at leading order} \label{sec:0thGO}

At the lowest order in $\hbar$, the Euler-Lagrange equations \eqref{eq:HJ_full} reduce to
\begin{subequations}

\begin{align}
    D \psi_0 &= 0, \label{eq:HJ1_h0}\\
    \bar{\psi}_0 D &= 0, \label{eq:HJ2_h0}\\
    \nabla_\mu {j_0}^\mu &= 0, \label{eq:transp_h0}
\end{align}
\end{subequations}
where we introduced the notation ${j_0}^\mu = \bar{\psi}_0 \gamma^\mu \psi_0$ for the conserved Dirac current at the lowest order in $\hbar$. Using Eqs. \eqref{eq:HJ1_h0} and \eqref{eq:HJ2_h0}, we can obtain an alternative expression for ${j_0}^\mu$. We start by writing the following identities:
\begin{subequations}
\begin{align}
    \Bar{\psi}_0 \gamma^\mu \left( m \psi_0 \right) &= - k_\nu \Bar{\psi}_0 \gamma^\mu \gamma^\nu \psi_0, \\
    \left( \Bar{\psi}_0 m \right) \gamma^\mu \psi_0 &= - k_\nu \Bar{\psi}_0 \gamma^\nu \gamma^\mu \psi_0. 
\end{align}
\end{subequations}
Adding these two equations and using the anticommutation property of gamma matrices, $\gamma^\mu \gamma^\nu + \gamma^\nu \gamma^\mu = - 2 g^{\mu \nu}$, we obtain
\begin{equation}
\begin{split}
    {j_0}^\mu = \Bar{\psi}_0 \gamma^\mu \psi_0 =\frac{1}{m} \mathcal{I}_0 k^\mu,
\end{split}
\end{equation}
where we defined the lowest-order intensity as $\mathcal{I}_0 = \Bar{\psi}_0 \psi_0$. The transport equation \eqref{eq:transp_h0} can be rewritten as
\begin{equation} \label{eq:transp_i0}
    \nabla_\mu \left( \mathcal{I}_0 k^\mu \right) = 0.
\end{equation}
Using Eqs. \eqref{eq:HJ1_h0} and \eqref{eq:HJ2_h0}, we can write
\begin{equation}
\begin{split}
    0 = - \bar{\psi}_0 D \psi_0 = \mathcal{I}_0 \left( \frac{1}{m} k_\mu k^\mu + m \right).
\end{split}
\end{equation}
Thus, we obtained the dispersion relation
\begin{equation} \label{eq:disp_0}
    k_\mu k^\mu = -m^2.
\end{equation}
It should be noted that
\begin{equation}
    2 \nabla_{[\nu} k_{\mu]} = - e F_{\mu \nu}.
\end{equation}
Thus, we can differentiate the dispersion relation \eqref{eq:disp_0} to obtain the Lorentz force law
\begin{equation}
    k^\mu \nabla_\mu k_\nu = e k^\mu F_{\mu \nu}.
\end{equation}

Equations \eqref{eq:HJ1_h0} and \eqref{eq:HJ2_h0} are homogeneous systems of linear algebraic equations for the unknowns $\psi_0$ and $\bar{\psi}_0$ \citep{Rubinow-Keller, audretsch,rudiger}. For these systems to admit nontrivial solutions, the determinant of the matrix $D$ must be zero. This condition is equivalent to the dispersion relation \eqref{eq:disp_0}:
\begin{equation}
    \det( D ) = 0 \quad \Leftrightarrow \quad k_\mu k^\mu = - m^2.  
\end{equation}
Under the restriction $k_\mu k^\mu = - m^2$, the matrix $D$ has rank 2. We can introduce a $4$-spinor basis $\{ \Sigma_0, \Sigma_1, \Pi_0, \Pi_1 \}$, where $\Sigma_0$ and $\Sigma_1$ are eigenspinors of $D$ with zero eigenvalue and $\Pi_0$ and $\Pi_1$ are eigenspinors of $D$, with eigenvalue $-2m$:
\begin{subequations} \label{eq:massive_eigenspinors}
\begin{alignat}{3}
    D \Sigma_A &= 0,   &&\bar{\Sigma}_A D = 0, \label{eq:eigenspinor1}\\
    D \Pi_A &= -2m \Pi_A, \qquad  &&\bar{\Pi}_A D =-2m \bar{\Pi}_A, \label{eq:eigenspinor2}
\end{alignat}
\end{subequations}
where $A, B = 0, 1$. Furthermore, the $4$-spinors satisfy the orthogonality relations
\begin{equation}
    \bar{\Sigma}_A \Sigma_B = -\bar{\Pi}_A \Pi_B = \delta_{A B},
\end{equation}
and the resolution of identity
\begin{equation} \label{eq:Res_id}
    \Sigma_A \bar{\Sigma}_A - \Pi_A \bar{\Pi}_A = \mathbb{I}_4.
\end{equation}
Here and in the following, we use an additional summation convention over repeated capital indices:
\begin{equation}
    \Sigma_A \bar{\Sigma}_A = \sum_{A = 0}^1 (\Sigma_A \bar{\Sigma}_A) = \Sigma_0 \bar{\Sigma}_0 + \Sigma_1 \bar{\Sigma}_1.
\end{equation}
Thus, Eqs. \eqref{eq:HJ1_h0} and \eqref{eq:HJ2_h0} are satisfied if the amplitude $\psi_0$ is an eigenspinor of $D$, with eigenvalue zero. The most general form for $\psi_0$ is
\begin{equation} \label{eq:amplitude_basis}
\begin{split}
    \psi_0 (x, k) &= \sqrt{\mathcal{I}_0(x)} \left[ z_0(x) \Sigma_0(x, k) + z_1(x) \Sigma_1(x, k) \right] \\
    &= \sqrt{\mathcal{I}_0(x)} z_A \Sigma_A,
\end{split}
\end{equation}
where $z_0$ and $z_1$ are scalar coefficients, satisfying the constraint
\begin{equation}
    \bar{z}_0 z_0 + \bar{z}_1 z_1 = \bar{z}_A z_A = 1.
\end{equation}
Note that since the matrix $D$ explicitly depends on $k_\mu$, its eigenspinors will, in general, also depend on $k_\mu$. As mentioned in Sec.~\ref{sec:wkb}, this can be viewed as a consequence of the amplitude $\psi$ being defined on the Lagrangian submanifold determined by $k_\mu$.

\subsection{Ray equations} \label{sec:ray0}

Equations \eqref{eq:transp_i0} and \eqref{eq:disp_0} represent a system of coupled partial differential equations
\begin{subequations}
\begin{align}
   \frac{1}{2}g^{\alpha \beta} \left(\nabla_\alpha S +e A_\alpha \right) \left(\nabla_\beta S +e A_\beta \right) &= -\frac{m^2}{2}, \\
    \nabla_\alpha \left[ \mathcal{I}_0 \left(\nabla^\alpha S +e A^\alpha \right) \right] &= 0, \label{eq:transp_HJ_0}
\end{align}
\end{subequations}
where the unknowns are $S$ and $\mathcal{I}_0$. The first equation is a Hamilton-Jacobi equation for the phase function $S$, and the second equation is a transport equation for the intensity $\mathcal{I}_0$. The Hamilton-Jacobi equation can be solved using the method of characteristics \citep[Sec. 46]{Arnold_book}. This is done by defining a Hamiltonian function $H$ on $T^*M$, such that
\begin{equation} \label{eq:HJ_0}
    H \left(x, \nabla S \right) = \frac{1}{2}g^{\alpha \beta} \left(\nabla_\alpha S +e A_\alpha \right) \left(\nabla_\beta S +e A_\beta \right) = -\frac{m^2}{2}.
\end{equation}
In this case, the Hamiltonian is
\begin{equation} \label{eq:H_0}
    H(x, p) = \frac{1}{2}g^{\alpha \beta} \left(p_\alpha +e A_\alpha \right) \left(p_\beta +e A_\beta \right),
\end{equation}
and Hamilton's equations (assuming the standard symplectic form on $T^*M$, $\Omega = d x^\mu \wedge d p_\mu$) take the following form:
\begin{subequations} \label{eq:EOM_0_xp}
\begin{align} 
    \begin{split} \dot{x}^\mu = \frac{\partial H}{\partial p_\mu} = p^\mu +e A^\mu , \end{split} \label{eq:EOM_0_x}\\
    \begin{split} \dot{p}_\mu = -\frac{\partial H}{\partial x^\mu} =  &-\frac{\partial_\mu g^{\alpha \beta}}{2}  \left(p_\alpha +e A_\alpha \right) \left(p_\beta +e A_\beta \right) \\
    & \quad - e \left(p^\alpha +e A^\alpha \right) \partial_\mu A_{\alpha}. \end{split} \label{eq:EOM_0_p}
\end{align}
\end{subequations}
Introducing a new momentum variable $k_\alpha = p_\alpha + e A_\alpha$, Hamilton's equations can be written in a more compact form:
\begin{subequations} \label{eq:EOM_0_XP}
\begin{align}
    \dot{x}^\mu &= k^\mu, \label{eq:EOM_0_X}\\
    \dot{k}_\mu &= -\frac{\partial_\mu g^{\alpha \beta}}{2} k_\alpha k_\beta + e k^\alpha F_{\alpha \mu}. \label{eq:EOM_0_V}
\end{align}
\end{subequations}
Given a solution $\{x^\mu(\tau), p_\mu(\tau)\}$ for Hamilton's equations, we can obtain a solution of the Hamilton-Jacobi equation \eqref{eq:HJ_0} by taking the phase function $S$ as follows \citep{goldstein}:
\begin{equation}
    S(x^\mu(\tau_1), p_\mu(\tau_1)) = \int_{\tau_0}^{\tau_1} d \tau L(x, \dot{x}, p, \dot{p})  + \text{const},
\end{equation}
where
\begin{equation} \label{eq:L_0}
    L(x, \dot{x}, p, \dot{p}) = \dot{x}^\mu p_\mu - H(x, p)
\end{equation}
is the corresponding Lagrangian. The ray equations \eqref{eq:EOM_0_xp} can also be obtained as the Euler-Lagrange equations corresponding to the Lagrangian $L$.  

Once the Hamilton-Jacobi equation is solved, the transport equation \eqref{eq:transp_i0} can also be analyzed (see Ref.~\citep{HJ_transport}). However, our main interest is in the ray equations governed by the Hamiltonian \eqref{eq:H_0} or by the Lagrangian \eqref{eq:L_0}. The ray equations \eqref{eq:EOM_0_XP} describe the timelike worldlines of massive charged particles. These equations can easily be rewritten as
\begin{equation} \label{eq:Lorentz_force1}
    \ddot{x}^\mu + \Gamma^\mu_{\alpha \beta} \dot{x}^\alpha \dot{x}^\beta - e \dot{x}^\alpha F\indices{_\alpha^\mu} = 0,
\end{equation}
or in the explicitly covariant form
\begin{equation} \label{eq:Lorentz_force2}
    k^\alpha \nabla_\alpha k^\mu = \dot{x}^\alpha \nabla_\alpha \dot{x}^\mu = e \dot{x}^\alpha F\indices{_\alpha^\mu}. 
\end{equation}

\subsection{WKB equations at next-to-leading order} \label{sec:WKB_1_massive}

Taking the Euler-Lagrange equations \eqref{eq:HJ1_fullA} and \eqref{eq:HJ2_fullA} at order $\hbar^1$ only, we obtain
\begin{subequations}\label{eq:psi1}
\begin{align} 
    D \psi_1  &= - i  \gamma^\mu \nabla_\mu \psi_0, \\
    \Bar{\psi}_1 D  &=  i \nabla_\mu \Bar{\psi}_0 \gamma^\mu. 
\end{align}
\end{subequations}
Given $\psi_0$, these are two inhomogeneous systems of linear algebraic equations, where the unknowns are $\psi_1$ and $\Bar{\psi}_1$. For any inhomogeneous system, the general solution can be written as the sum of the solution for the homogeneous system, and a particular solution for the inhomogeneous system. We can write $\psi_1$ as
\begin{equation}
    \psi_1 = b_0 \Sigma_0 + b_1 \Sigma_1 + \psi_p,
\end{equation}
where $b_{0, 1}$ are scalar coefficients and $\psi_p$ is a particular solution of the inhomogeneous system. The system will admit nontrivial solutions if and only if the right-hand side of the inhomogeneous equation is orthogonal to all the solutions of the transposed homogeneous equation. Such solutions are always a linear combination of $\Sigma_0$ and $\Sigma_1$. Therefore, we have the following solvability conditions \citep{Rubinow-Keller, audretsch, rudiger}, which impose additional constraints on $\psi_0$ and $\bar{\psi}_0$:
\begin{subequations}
\begin{align}
    \bar{\Sigma}_0 \gamma^\mu \nabla_\mu \psi_0 = \bar{\Sigma}_1 \gamma^\mu \nabla_\mu \psi_0 = 0, \\
    \nabla_\mu \Bar{\psi}_0 \gamma^\mu \Sigma_0 = \nabla_\mu \Bar{\psi}_0 \gamma^\mu \Sigma_1 = 0.
\end{align}
\end{subequations}
We can rewrite these equations using the expansion of $\psi_0$ given in Eq.~\eqref{eq:amplitude_basis}, and the transport equation \eqref{eq:transp_i0}:
\begin{subequations}
\begin{align}
    k^\mu \nabla_\mu z_A &= \frac{1}{2} z_A \nabla_\mu k^\mu - m \bar{\Sigma}_A \gamma^\mu \nabla_\mu \Sigma_B z_B, \\
    k^\mu \nabla_\mu \bar{z}_B &= \frac{1}{2} \bar{z}_B \nabla_\mu k^\mu - m \bar{z}_A \bar{\Sigma}_A \gamma^\mu \nabla_\mu \Sigma_B.
\end{align}
\end{subequations}
Using the identity
\begin{equation}
    \bar{\Sigma}_A \gamma^\mu \Sigma_B = \frac{1}{m} \delta_{A B} k^\mu,
\end{equation}
and its derivative 
\begin{equation}
    \delta_{A B} \nabla_\mu k^\mu = m \left( \nabla_\mu \bar{\Sigma}_A \gamma^\mu \Sigma_B + \bar{\Sigma}_A \gamma^\mu \nabla_\mu \Sigma_B \right), 
\end{equation}
we obtain
\begin{subequations} \label{eq:zdot1}
\begin{align}
    k^\mu \nabla_\mu z_A &= i M_{A B} z_B, \\
    k^\mu \nabla_\mu \bar{z}_B &= -i \bar{z}_A M_{A B},
\end{align}   
\end{subequations}
where the $2 \times 2$ Hermitian matrix $M$ has components
\begin{equation}
    M_{A B} = \frac{i m}{2} \left( \bar{\Sigma}_A \gamma^\mu \nabla_\mu \Sigma_B - \nabla_\mu \bar{\Sigma}_A \gamma^\mu \Sigma_B \right).    
\end{equation}
Using the properties of the eigenspinors given in Eqs. \eqref{eq:massive_eigenspinors}--\eqref{eq:Res_id}, the matrix $M_{A B}$ can be rewritten as
\begin{equation}
    M_{A B} = \frac{i}{2} k^\mu \left( \bar{\Sigma}_A \nabla_\mu \Sigma_B - \nabla_\mu \bar{\Sigma}_A \Sigma_B \right) - \frac{e}{4} F_{\mu \nu} \bar{\Sigma}_A \sigma^{\mu \nu} \Sigma_B .
\end{equation}
Here, the first term represents the Berry connection, and the second term represents the so-called ``no-name" term. The no-name term was first introduced in a general context by Littlejohn and Flynn \citep{Littlejohn1991} (see also Ref. \citep{Emmrich1996} for a geometric discussion), and its role in the WKB approximation to the Dirac equation was discussed in Ref.~\citep{Stone2015(2)}.

We can write Eq.~\eqref{eq:zdot1} in a more compact form by introducing the following two-dimensional unit complex vectors:
\begin{equation}
    z = \begin{pmatrix} z_0 \\ z_1 \end{pmatrix}, \qquad \qquad \bar{z} = \begin{pmatrix} \bar{z}_0 & \bar{z}_1 \end{pmatrix}.
\end{equation}
We also introduce the following notation:
\begin{subequations}
\begin{align}
    B_{\mu A B} (x, k) &= \frac{i}{2} \left( \bar{\Sigma}_A \nabla_\mu \Sigma_B - \nabla_\mu \bar{\Sigma}_A \Sigma_B \right), \\
    s\indices{^{\mu \nu}_{A B}} (x, k) &= \frac{1}{2} \bar{\Sigma}_A \sigma^{\mu \nu} \Sigma_B,
\end{align}
\end{subequations}
where $\sigma^{\mu\nu} = \frac{i}{2}[\gamma^\mu, \gamma^\nu]$. The Berry connection $B_{\mu}$ is a $2 \times 2$ matrix-valued one-form, while $s^{\mu \nu}$ is a $2 \times 2$ matrix-valued antisymmetric tensor. Depending on the context, we will sometimes omit the matrix indices $A, B$. Then, Eq.~\eqref{eq:zdot1} can be written as
\begin{subequations}
\begin{align}
    k^\mu \nabla_\mu z &= i \left( k^\mu B_\mu - \frac{e}{2} F_{\mu \nu} s^{\mu \nu}  \right) z, \\
    k^\mu \nabla_\mu \bar{z} &= -i \bar{z} \left( k^\mu B_\mu - \frac{e}{2} F_{\mu \nu} s^{\mu \nu} \right).
\end{align}   
\end{subequations}
If we restrict $z$ to a worldline $x^\mu(\tau)$, which is a solution of the ray equations \eqref{eq:EOM_0_XP}, we can write
\begin{subequations} \label{eq:transp_zdot}
\begin{align}
    \dot{z} &= i \left( k^\mu B_\mu - \frac{e}{2} F_{\mu \nu} s^{\mu \nu}  \right) z, \\
    \dot{\bar{z}} &= -i \bar{z} \left( k^\mu B_\mu - \frac{e}{2} F_{\mu \nu} s^{\mu \nu} \right).
\end{align}   
\end{subequations}
These equations describe the evolution of the spin degree of freedom along the worldline $x^\mu(\tau)$.

It is important to emphasize how the covariant derivatives act on the eigenspinors $\Sigma_A$, which are defined on the Lagrangian submanifold. Applying the chain rule and using the horizontal and vertical derivatives defined in Appendix \ref{app:derivative_TM}, we obtain
\begin{equation} \label{eq:chain_rule}
\begin{split}
    k^\mu \nabla_\mu \Sigma_A &= k^\mu \nabla_\mu \left[ \Sigma_A (x, k) \right] \\
    &= k^\mu \left( \hnabla{}_\mu \Sigma_A \right) (x, k) \\
    & \qquad+ k^\mu \left( \nabla_\mu k_\nu \right) \left(\vnabla{}^\nu \Sigma_A \right) (x, k) \\
    &= k^\mu \hnabla{}_\mu \Sigma_A + e k^\mu F_{\mu \nu} \vnabla{}^\nu \Sigma_A.
\end{split}
\end{equation}
The expression of the Berry connection becomes
\begin{equation} \label{eq:berry_conn}
\begin{split}
    k^\mu B_{\mu A B} &= \frac{i}{2} k^\mu \left( \bar{\Sigma}_A \hnabla{}_\mu \Sigma_B - \hnabla{}_\mu \bar{\Sigma}_A \Sigma_B \right) \\
    & \qquad+ \frac{i e}{2} k^\mu F_{\mu \nu} \left( \bar{\Sigma}_A \vnabla{}^\nu \Sigma_B - \vnabla{}^\nu \bar{\Sigma}_A \Sigma_B \right).
\end{split}
\end{equation}

\subsection{Geometric definition of the Berry connection and Berry curvature} \label{sec:geometry_Berry}

A general discussion about the geometry of the transport equation arising from the WKB approximation of multicomponent wave equations can be found in Ref. \citep{Emmrich1996} (see also Refs.~\citep{Littlejohn1991,Emmrich1998,BOLTE1999}). Here, we specialize this discussion to the case of the Dirac equation, focusing on the geometry of the Berry connection and the corresponding Berry curvature.

The WKB approximation of multicomponent wave equations generally results in a Hamilton-Jacobi equation for the phase $S$ and a transport equation for the amplitude $\psi_0$. In the present case, the Hamilton-Jacobi equation for $S$ was discussed in Sec. \ref{sec:ray0}, and the transport equation for $\psi_0$ is divided into two parts. The first one describes the evolution of the intensity $\mathcal{I}_0$ and is given in Eq. \eqref{eq:transp_i0}, while the second one describes the evolution of the spin degree of freedom, as presented in Eq. \eqref{eq:transp_zdot}.

Since the amplitude $\psi_0$ is defined on the Lagrangian submanifold, the Berry connection must be defined as a connection on an appropriate 4-spinor bundle with base space the Lagrangian submanifold. Furthermore, the amplitude $\psi_0$ is an eigenspinor of $D$ with eigenvalue $\lambda = 0$, so the appropriate bundle is then the $\lambda$-eigenbundle of the 4-spinor bundle with base space the Lagrangian submanifold. Then, the Berry connection has to be a Lie algebra-valued one-form defined on the Lagrangian submanifold. This is clearly not the case for $B_\mu$, which is written as a Lie algebra-valued one-form on spacetime. 

A connection one-form defined on the Lagrangian submanifold should be contracted with a tangent vector to the Lagrangian submanifold. By definition, the tangent vectors to the Lagrangian submanifold are the Hamiltonian vector fields. Working in the $(x, k)$ coordinates, the Hamiltonian vector field corresponding to the ray equations \eqref{eq:EOM_0_XP} is
\begin{equation} \label{eq:X_H}
    X_H = k^\mu \frac{\partial}{\partial x^\mu} + \left( \Gamma^\rho_{\nu\mu} k_\rho k^\nu + e k^\nu F_{\nu \mu} \right) \frac{\partial}{\partial k_\mu}
\end{equation}
We can obtain the appropriate definition of the Berry connection as follows. Using the definition of the horizontal derivative, we can rewrite Eq. \eqref{eq:berry_conn} as
\begin{equation}
\begin{split}
    &\mathcal{B}_{A B}(X_H) = k^\mu B_{\mu A B} = \frac{i}{2} k^\mu \left( \bar{\Sigma}_A \nabla_\mu \Sigma_B - \nabla_\mu \bar{\Sigma}_A \Sigma_B \right) \\
    &\quad+ \frac{i}{2}\left( \Gamma^\rho_{\nu\mu} k_\rho k^\nu + e k^\nu F_{\nu \mu} \right) \left( \bar{\Sigma}_A \vnabla{}^\mu \Sigma_B - \vnabla{}^\mu \bar{\Sigma}_A \Sigma_B \right),
\end{split}
\end{equation}
and we obtain
\begin{equation}
\begin{split}
    \mathcal{B}_{A B} &= \frac{i}{2} \left( \bar{\Sigma}_A \nabla_\mu \Sigma_B - \nabla_\mu \bar{\Sigma}_A \Sigma_B \right) d x^\mu \\
    &\qquad+ \frac{i}{2} \left( \bar{\Sigma}_A \vnabla{}^\mu \Sigma_B - \vnabla{}^\mu \bar{\Sigma}_A \Sigma_B \right) d k_\mu.
\end{split}
\end{equation}
This is the appropriately defined Berry connection that we were looking for, which is a Lie algebra-valued one-form defined on the Lagrangian submanifold. The corresponding Lie algebra is $\mathfrak{u}(2)$, since the one-form $i \mathcal{B}$ takes values in the space of two-dimensional anti-Hermitian matrices.

The curvature of the connection $\mathcal{B}$ can be calculated using the standard definition \citep[Sec. 2.3.2]{Berry_book}
\begin{equation} \label{eq:def_curvature}
    \mathcal{F} = d \mathcal{B} - i[\mathcal{B}, \mathcal{B}].
\end{equation}
This is the Berry curvature, and it plays an important role in the spin Hall effect correction to the ray equations, as we will discuss in the next sections. In coordinates, the expression of the Berry curvature is
\begin{equation}
\begin{split}
    \mathcal{F} &= (\mathcal{F}_{x x})_{\mu \nu} dx^\mu dx^\nu + (\mathcal{F}_{k x})\indices{_\mu^\nu} dx^\mu dk_\nu \\
    &\qquad + (\mathcal{F}_{x k})\indices{^\mu_\nu} dk_\mu dx^\nu + (\mathcal{F}_{k k})_{\mu \nu} dk^\mu dk^\nu, 
\end{split}
\end{equation}
where
\begin{align}
    \left( \mathcal{F}_{x x} \right)_{\mu \nu} &= \frac{\partial \left( \mathcal{B}_x \right)_\nu}{\partial x^\mu} - \frac{\partial \left( \mathcal{B}_x \right)_\mu}{\partial x^\nu} - i \left[ (\mathcal{B}_x)_\mu, (\mathcal{B}_x)_\nu \right], \\
    \left( \mathcal{F}_{k k} \right)^{\mu \nu} &= \frac{\partial \left( \mathcal{B}_k \right)^\nu }{\partial k_\mu} - \frac{\partial \left( \mathcal{B}_k \right)^\mu }{\partial k_\nu} - i \left[ (\mathcal{B}_k)^\mu, (\mathcal{B}_k)^\nu \right], \\
    \left( \mathcal{F}_{k x} \right)\indices{_\mu^\nu} &= \frac{\partial \left( \mathcal{B}_k \right)^\nu }{\partial x^\mu} - \frac{\partial \left( \mathcal{B}_x \right)_\mu }{\partial k_\nu} - i \left[ (\mathcal{B}_x)_\mu, (\mathcal{B}_k)^\nu \right], \\
    \left( \mathcal{F}_{x k} \right)\indices{^\nu_\mu} &= - \left( \mathcal{F}_{k x} \right)\indices{_\mu^\nu},
\end{align}
and
\begin{subequations}
\begin{align}
    \left( \mathcal{B}_{x} \right)_{\mu A B} &= \frac{i}{2} \left( \bar{\Sigma}_A \nabla_\mu \Sigma_B - \nabla_\mu \bar{\Sigma}_A \Sigma_B \right), \\
    \left( \mathcal{B}_{k} \right)\indices{^\mu_{A B}} &= \frac{i}{2} \left( \bar{\Sigma}_A \vnabla{}^\mu \Sigma_B - \vnabla{}^\mu \bar{\Sigma}_A \Sigma_B \right).
\end{align}
\end{subequations}
Using the properties of the eigenspinors given in Eqs. \eqref{eq:eigenspinor1}--\eqref{eq:Res_id}, the components of the Berry curvature can be explicitly computed, as shown in Appendix \ref{app:curvature}:
\begin{subequations} \label{eq:Berry_comp}
\begin{align}
    \left( \mathcal{F}_{x x} \right)_{\mu \nu} &= -\frac{1}{2} R_{\mu \nu \alpha \beta} s^{\alpha \beta} + \frac{1}{m^2} k_\rho k_\sigma \Gamma^\rho_{\alpha \mu} \Gamma^\sigma_{\beta \nu} s^{\alpha \beta}, \label{eq:Fxx}\\
    \left( \mathcal{F}_{k k} \right)^{\mu \nu} &= \frac{1}{m^2} s^{\mu \nu}, \\
    \left( \mathcal{F}_{k x} \right)\indices{_\mu^\nu} &= - \left( \mathcal{F}_{x k} \right)\indices{^\nu_\mu} = - \frac{1}{m^2} k_\rho \Gamma^\rho_{\mu \alpha} s^{\alpha \nu}. \label{eq:Fpx}
\end{align}
\end{subequations}
The components of the Berry curvature were also calculated in Ref. \citep{Stone2015(2)}, although only for the case of Minkowski spacetime. Restricting to Minkowski spacetime, the only nonzero component of the Berry curvature is $\left( \mathcal{F}_{k k} \right)^{\mu \nu}$, and in this case our result agrees with the result presented in Ref. \citep[Eq. 30]{Stone2015(2)}.

\subsection{Effective dispersion relation and spin-orbit coupling} \label{sec:eff_disp}

In the standard WKB treatment, the equations for each individual order in $\hbar$ are set to zero. The resulting ray equations \eqref{eq:EOM_0_XP} are used to determine the transport of the spin degree of freedom, by Eq.~\eqref{eq:transp_zdot}. However, with this approach there is no backreaction from the spin degree of freedom, described by $z(\tau)$, on the orbital degrees of freedom, described by the rays $x^\mu(\tau)$ and $k(\tau)$. To properly take into account the spin-orbit coupling between the spin dynamics and the ray dynamics, we derive here an effective dispersion relation, containing $\mathcal{O}(\hbar^1)$ corrections to the dispersion relation given in Eq. \eqref{eq:disp_0}. This represents a weaker condition compared to the standard WKB treatment, where terms of different orders in $\hbar$ are set to zero individually. Instead, here we only require that the combined sum of the terms of order $\hbar^0$ and $\hbar^1$ vanishes. The effective dispersion relation is obtained by taking the Euler-Lagrange equations \eqref{eq:HJ_full}, but without treating terms of different orders in $\hbar$ separately. The effective dispersion relation is then treated as an effective Hamilton-Jacobi equation, and the resulting ray equations contain spin-dependent correction terms, describing the gravitational spin Hall effect of Dirac particles.

Starting with Eq. \eqref{eq:HJ_full}, we can write
\begin{subequations}
\begin{align}
    \Bar{\psi} \left( \gamma^\mu k_\mu + m \right) \psi - i \hbar \Bar{\psi} \gamma^\mu \nabla_\mu \psi &= \mathcal{O}(\hbar^2), \\
    \Bar{\psi} \left( \gamma^\mu k_\mu + m \right) \psi + i \hbar \nabla_\mu \Bar{\psi} \gamma^\mu \psi &= \mathcal{O}(\hbar^2).
\end{align}
\end{subequations}
By adding these two equations, we obtain
\begin{equation} \label{eq:eff_disp_1}
    k_\mu \Bar{\psi} \gamma^\mu \psi + m \Bar{\psi} \psi - \frac{i \hbar}{2} \left( \Bar{\psi} \gamma^\mu \nabla_\mu \psi - \nabla_\mu \Bar{\psi} \gamma^\mu \psi \right) = \mathcal{O}(\hbar^2)
\end{equation}
The Dirac current $j^\mu = \Bar{\psi} \gamma^\mu \psi + \mathcal{O}(\hbar^2)$ can be rewritten in a different form. Using Eqs. \eqref{eq:HJ1_fullA} and \eqref{eq:HJ2_fullA}, we get
\begin{equation}
\begin{split}
    \Bar{\psi} \gamma^\mu \left( m \psi \right) &= \Bar{\psi} \gamma^\mu \left( -\gamma^\nu k_\nu \psi + i \hbar \gamma^\nu \nabla_\nu \psi \right) + \mathcal{O}(\hbar^2) \\
    \left( \Bar{\psi} m \right) \gamma^\mu \psi &= \left( -\Bar{\psi} \gamma^\nu k_\nu - i \hbar \nabla_\nu \Bar{\psi} \gamma^\nu \right) \gamma^\mu \psi + \mathcal{O}(\hbar^2)\\
\end{split}
\end{equation}
Adding the above equations, we obtain
\begin{equation} \label{eq:j1}
\begin{split}
    j^\mu &= - \frac{1}{2m} k_\nu \Bar{\psi} \left( \gamma^\mu \gamma^\nu + \gamma^\nu \gamma^\mu \right) \psi \\
    & + \frac{i \hbar}{2m} \left( \Bar{\psi} \gamma^\mu \gamma^\nu \nabla_\nu \psi - \nabla_\nu \Bar{\psi} \gamma^\nu \gamma^\mu \psi \right) + \mathcal{O}(\hbar^2)\\
    &= \frac{1}{m} k^\mu \Bar{\psi} \psi - \frac{i\hbar}{2m} g^{\mu \nu} \left( \Bar{\psi} \nabla_\nu \psi - \nabla_\nu \Bar{\psi} \psi \right) \\
    &\qquad + \frac{\hbar}{2m} \nabla_\nu \left( \Bar{\psi} \sigma^{\mu \nu} \psi \right) + \mathcal{O}(\hbar^2).
\end{split}
\end{equation}
Using the above result, Eq. \eqref{eq:eff_disp_1} can be rewritten as
\begin{equation}
\begin{split}
    &\bar{\psi} \psi k_\mu k^\mu - \frac{i \hbar}{2} k^\mu \left( \Bar{\psi}_0  \nabla_\mu \psi_0 - \nabla_\mu \Bar{\psi}_0 \psi_0 \right) \\
    &+ \frac{\hbar}{2} k_\mu \nabla_\nu \left( \Bar{\psi}_0 \sigma^{\mu \nu} \psi_0 \right) - \frac{i \hbar m}{2} \left( \Bar{\psi}_0 \gamma^\mu \nabla_\mu \psi_0 - \nabla_\mu \Bar{\psi}_0 \gamma^\mu \psi_0 \right) \\
    & \qquad = -m^2 \bar{\psi} \psi + \mathcal{O}(\hbar^2).
\end{split}
\end{equation}
We expand the $\mathcal{O}(\hbar)$ terms above using Eq. \eqref{eq:amplitude_basis}:
\begin{equation}
\begin{split}
    &\frac{i \hbar}{2} k^\mu \left( \Bar{\psi}_0  \nabla_\mu \psi_0 - \nabla_\mu \Bar{\psi}_0  \psi_0 \right) \\
    &\quad= \frac{i \hbar \mathcal{I}_0}{2} k^\mu \left( z^\dagger \nabla_\mu z - \nabla_\mu z^\dagger z \right)+ \hbar \mathcal{I}_0 k^\mu z^\dagger B_\mu z.  
\end{split}
\end{equation}
Using Eqs. \eqref{eq:HJ1_h0} and \eqref{eq:HJ2_h0}, we can write
\begin{equation}
\begin{split}
    &k_\mu \nabla_\nu \left( \Bar{\psi}_0 \sigma^{\mu \nu} \psi_0 \right) = e \mathcal{I}_0 F_{\mu \nu} z^\dagger s^{\mu \nu} z.
\end{split}
\end{equation}
We also have
\begin{equation}
\begin{split}
    &\frac{i \hbar}{2}\left( \Bar{\psi}_0 \gamma^\mu \nabla_\mu \psi_0 - \nabla_\mu \Bar{\psi}_0 \gamma^\mu \psi_0 \right) \\
    &\quad= \frac{i \hbar \mathcal{I}_0}{2 m}  k^\mu \left( z^\dagger \nabla_\mu z - \nabla_\mu z^\dagger z \right) \\
    &\qquad + \frac{ \hbar \mathcal{I}_0}{m} k^\mu z^\dagger B_\mu z - \frac{e \hbar \mathcal{I}_0}{2 m} F_{\mu \nu} z^\dagger s^{\mu \nu} z.
\end{split}
\end{equation}
At this stage, it is also convenient to introduce the spin tensor 
\begin{equation} \label{eq:Spin_tensor_massive}
    S^{\mu \nu} = \hbar \bar{z} s^{\mu \nu} z = \frac{\hbar}{2} \frac{\Bar{\psi}_0 \sigma^{\mu \nu} \psi_0}{\Bar{\psi}_0 \psi_0}.
\end{equation}
The role of the spin tensor is to encode the angular momentum carried by the wave packet, and similar definitions have also been considered in Refs. \citep{audretsch,rudiger,Stone2015(2)}. Using Eq. \eqref{eq:massive_eigenspinors}, it is straightforward to show that the spin tensor is orthogonal to the momentum:
\begin{equation}
    S^{\mu \nu} k_\nu = 0.
\end{equation}
Combining the above equations, we obtain the effective dispersion relation
\begin{equation} \label{eq:disp_1}
\begin{split}
    &\frac{1}{2} k_\mu k^\mu - \frac{i \hbar}{2} k^\mu \left( \bar{z} \nabla_\mu z - \nabla_\mu \bar{z} z \right) \\
    & \qquad- \hbar k^\mu \bar{z} B_\mu z + \frac{e}{2} F_{\mu \nu} S^{\mu \nu} = -\frac{m^2}{2} + \mathcal{O}(\hbar^2).
\end{split}
\end{equation}
The above dispersion relation contains $\mathcal{O}(\hbar)$ corrections to the standard dispersion relation obtained in Eq. \eqref{eq:disp_0}. These additional terms describe the spin-orbit coupling.

\subsubsection{Effective ray equations}\label{sec:effectivemotion}

Following the same method as in Refs. \citep{GSHE2020,GSHE_GW}, we now derive effective ray equations that contain spin-dependent correction terms. These equations are meant to describe the gravitational spin Hall effect of Dirac wave packets. We start with the effective dispersion relation \eqref{eq:disp_1} and treat it as an effective Hamilton-Jacobi equation for the phase function $S$:
\begin{equation}
\begin{split}
    &\frac{1}{2}g^{\alpha \beta} \left(\nabla_\alpha S +e A_\alpha \right) \left(\nabla_\beta S +e A_\beta \right) - \frac{i \hbar}{2} \left( \bar{z} \dot{z} - \dot{\bar{z}} z \right) \\
    &- \hbar \left(\nabla^\alpha S +e A^\alpha \right) \bar{z} B_\alpha z + \frac{e}{2} F_{\alpha \beta} S^{\alpha \beta} = -\frac{m^2}{2} + \mathcal{O}(\hbar^2).
\end{split}
\end{equation}
We define the corresponding Hamiltonian function
\begin{equation}
\begin{split}
    H &= \frac{1}{2}g^{\alpha \beta} \left(p_\alpha +e A_\alpha \right) \left(p_\beta +e A_\beta \right) - \frac{i \hbar}{2} \left( \bar{z} \dot{z} - \dot{\bar{z}} z \right) \\
    &\qquad -\hbar \left(p^\alpha +e A^\alpha \right) \bar{z} B_\alpha z + \frac{e}{2} F_{\alpha \beta} S^{\alpha \beta},
\end{split}
\end{equation}
and we solve for the phase function $S$ as in Sec. \ref{sec:ray0}:
\begin{equation}
    S(x^\alpha(\tau), p_\alpha(\tau), z(\tau), \bar{z}(\tau)) = \int_{\tau_0}^{\tau} d \tau' L  + \text{const},
\end{equation}
where the Lagrangian is
\begin{equation}
\begin{split}
    L = \dot{x}^\alpha p_\alpha &- \frac{1}{2}g^{\alpha \beta} \left(p_\alpha +e A_\alpha \right) \left(p_\beta +e A_\beta \right) + \frac{i \hbar}{2} \left( \bar{z} \dot{z} - \dot{\bar{z}} z \right) \\
    &+ \hbar \left(p^\alpha +e A^\alpha \right) \bar{z} B_\alpha z - \frac{e}{2} F_{\alpha \beta} S^{\alpha \beta}.
\end{split}
\end{equation}
Note that the Lagrangian is a scalar function defined on $T(T^*M \times \mathbb{C}^2)$, and the effective ray dynamics is given by the Euler-Lagrange equations
\begin{equation}
    \frac{\partial L}{\partial w} - \frac{d}{d \tau} \frac{\partial L}{\partial \dot{w}} = 0,
\end{equation}
where $w \in \{x^\mu, p_\mu, z, \bar{z} \}$. The Euler-Lagrange equations are
\begin{subequations} \label{eq:ray_eq_1}
\begin{align} 
    \dot{x}^\mu &= k^\mu - \hbar \bar{z} B^\mu z - \hbar k^\alpha \bar{z} \frac{\partial B_\alpha}{\partial p_\mu} z + \frac{e}{2} F_{\alpha \beta} \frac{\partial S^{\alpha \beta}}{\partial p_\mu} , \label{eq:dot_x}\\
    \begin{split}
    \dot{p}_\mu &= \Gamma^\alpha_{\beta \mu} k_\alpha k^\beta - e k^\alpha \partial_\mu A_{\alpha}  + \hbar k^\alpha \bar{z} ( \partial_\mu B_{\alpha}) z \\
    &\qquad+ e \hbar (\partial_\mu A^\alpha) \bar{z} B_\alpha z - \frac{e}{2} \partial_\mu ( F_{\alpha \beta} S^{\alpha \beta} ) ,
    \end{split} \\
    \dot z &= i \left( k^\mu B_\mu - \frac{e}{2} F_{\mu \nu} s^{\mu \nu}  \right) z, \\
    \dot{\bar{z}} &= -i \bar{z} \left( k^\mu B_\mu - \frac{e}{2} F_{\mu \nu} s^{\mu \nu}  \right). \label{eq:dot_zbar}
\end{align}
\end{subequations}
These equations contain spin-dependent correction terms of $\mathcal{O}(h^1)$ to the ray equations obtained in Eq. \eqref{eq:EOM_0_XP}. The $\mathcal{O}(h^1)$ terms reflect the spin-orbit coupling between the external and internal degrees of freedom, resulting in the gravitational spin Hall effect of localized Dirac wave packets.

\subsubsection{Noncanonical coordinates}

The effective ray equations \eqref{eq:ray_eq_1} can also be formulated as a Hamiltonian system on the symplectic manifold $T^*M \times \mathbb{C}^2$. Using canonical coordinates $(x, p, z, \bar{z})$, the corresponding Hamiltonian function is
\begin{equation}
\begin{split}
    &H(x, p, z, \bar{z}) = \frac{1}{2}g^{\alpha \beta} \left(p_\alpha +e A_\alpha \right) \left(p_\beta +e A_\beta \right) \\
    &\qquad\qquad -\hbar \left(p^\alpha +e A^\alpha \right) \bar{z} B_\alpha z + \frac{e}{2} F_{\alpha \beta} S^{\alpha \beta},
\end{split}
\end{equation}
and the symplectic two-form is 
\begin{equation}
    \Omega = dx^\alpha \wedge dp_\alpha + i \hbar dz \wedge d\bar{z}.
\end{equation}
In this symplectic setup, Hamilton's equations are \citep[Sec. 3.3]{abraham1978foundations}
\begin{equation} \label{eq:cov_Hamilton_eq}
    \Omega(X_H, \,\,\cdot\,\,) = dH,
\end{equation}
where the Hamiltonian vector field $X_H$ can be expressed in coordinates as
\begin{equation}
    X_H = \dot{x}^\mu \frac{\partial}{\partial x^\mu} + \dot{p}_\mu \frac{\partial}{\partial p_\mu} + \dot{z} \frac{\partial}{\partial z} + \dot{\bar{z}} \frac{\partial}{\partial \bar{z}}.
\end{equation}
By solving for the components of the Hamiltonian vector field, we obtain the effective ray equations \eqref{eq:ray_eq_1} in the following form:
\begin{subequations} \label{eq:GHSE_massive_2}
\begin{equation}
\begin{alignedat}{2}
    \dot{x}^\mu &= \frac{\partial H}{\partial p_\mu}, \qquad && \dot{p}_\mu = -\frac{\partial H}{\partial x^\mu}, \\
    \dot{z} &= -\frac{i}{\hbar} \frac{\partial H}{\partial \bar{z}},  \qquad  && \dot{\bar{z}} = \frac{i}{\hbar} \frac{\partial H}{\partial z}.
\end{alignedat}
\end{equation}
\end{subequations}
However, these ray equations are gauge-dependent. First of all, they depend on the electromagnetic gauge potential $A_\mu$. More importantly, the ray equations depend on the Berry connection $B_\mu$, which is gauge-dependent in the sense that it depends on the choice of eigenspinors $\Sigma_A$. Thus, we aim to remove both these gauge dependencies from the equations of motion by introducing noncanonical coordinates. As a first step, we rewrite the Hamiltonian, symplectic form, and effective ray equations in the new coordinate system $(x, k, z, \bar{z})$, where $k_\mu = p_\mu + e A_\mu$. In these coordinates, the Hamiltonian is
\begin{equation}
    H(x, k, z, \bar{z}) = \frac{1}{2}g^{\alpha \beta} k_\alpha k_\beta - \hbar k^\alpha \bar{z} B_\alpha z + \frac{e}{2} F_{\alpha \beta} S^{\alpha \beta}.
\end{equation}
Applying the standard coordinate transformation rules for two-forms, the symplectic form can be expressed in the new coordinates $(x, k, z, \bar{z})$ as
\begin{equation}
    \Omega = e F_{\alpha \beta} dx^\alpha dx^\beta + dx^\alpha \wedge dk_\alpha + i \hbar dz \wedge d\bar{z}.
\end{equation}
Using Eq. \eqref{eq:cov_Hamilton_eq}, we can obtain the effective ray equations as the components of the Hamiltonian vector field in the new coordinates:
\begin{equation}
\begin{alignedat}{2}
    \dot{x}^\mu &= \frac{\partial H}{\partial k_\mu}, \qquad && \dot{k}_\mu = -\frac{\partial H}{\partial x^\mu} + e \dot{x}^\nu F_{\nu \mu}, \\
    \dot{z} &= -\frac{i}{\hbar} \frac{\partial H}{\partial \bar{z}},  \qquad  && \dot{\bar{z}} = \frac{i}{\hbar} \frac{\partial H}{\partial z}.
\end{alignedat}
\end{equation}
These equations no longer depend on the electromagnetic gauge potential $A_\mu$. To eliminate the Berry connection from the Hamiltonian, we perform the following coordinate transformation:
\begin{subequations}
\begin{align}
    X^\mu &= x^\mu + \hbar \bar{z} (\mathcal{B}_k)^\mu z, \\
    P_\mu &= k_\mu - \hbar \bar{z} (\mathcal{B}_x)_\mu z - e \hbar F_{\mu \nu} \bar{z} (\mathcal{B}_k)^\nu z .
\end{align}
\end{subequations}
This type of coordinate transformation was first introduced in Ref. \citep{Littlejohn1991}. It has the effect of eliminating the gauge-dependent Berry connection from the Hamiltonian, at the cost of introducing an additional term in the symplectic form, represented by the gauge-independent Berry curvature. A similar coordinate transformation was used in Refs. \citep{GSHE2020,GSHE_GW}. The Hamiltonian function becomes
\begin{equation}
\begin{split}
    &H(x^\mu, k_\mu, z, \bar{z}) = H(X^\mu -  \hbar \bar{z} (\mathcal{B}_k)^\mu z,\\
    &\qquad \qquad P_\mu + \hbar \bar{z} (\mathcal{B}_x)_\mu z + e \hbar F_{\mu \nu} \bar{z} (\mathcal{B}_k)^\nu z, z, \bar{z}) \\
    &= H(X^\mu, P_\mu, z, \bar{z}) - \hbar \frac{\partial H}{\partial X^\mu} \bar{z} (\mathcal{B}_k)^\mu z \\
    &\qquad + \hbar \frac{\partial H}{\partial P_\mu} \left[ \bar{z} (\mathcal{B}_x)_\mu z + e F_{\mu \nu} \bar{z} (\mathcal{B}_k)^\nu z \right] + \mathcal{O}(\hbar^2).
\end{split}
\end{equation}
Thus, the Hamiltonian in the noncanonical coordinates $(X, P, z, \bar{z})$ is
\begin{equation}
    H(X, P, z, \bar{z}) = \frac{1}{2}g^{\alpha \beta} P_\alpha P_\beta + \frac{e}{2} F_{\alpha \beta} S^{\alpha \beta} + \mathcal{O}(\hbar^2).
\end{equation}
Applying the same coordinate transformation to the symplectic form, we obtain
\begin{equation}
\begin{split}
    \Omega &= e F_{\alpha \beta} dX^\alpha dX^\beta + dX^\alpha \wedge dP_\alpha + i \hbar dz \wedge d\bar{z} \\
    &\quad - \hbar \bar{z} \left[ \frac{\partial \left( \mathcal{B}_x \right)_\beta}{\partial X^\alpha} - \frac{\partial \left( \mathcal{B}_x \right)_\alpha}{\partial X^\beta} \right] z dX^\alpha dX^\beta \\
    &\quad - \hbar \bar{z} \left[ \frac{\partial \left( \mathcal{B}_k \right)^\beta }{\partial X^\alpha} - \frac{\partial \left( \mathcal{B}_x \right)_\alpha }{\partial P_\beta} \right] z dX^\alpha dP_\beta \\ 
    &\quad - \hbar \bar{z} \left[ \frac{\partial \left( \mathcal{B}_x \right)_\beta }{\partial P_\alpha} - \frac{\partial \left( \mathcal{B}_k \right)^\alpha }{\partial X^\beta} \right] z dP_\alpha dX^\beta \\
    &\quad - \hbar \bar{z} \left[ \frac{\partial \left( \mathcal{B}_k \right)^\beta }{\partial P_\alpha} - \frac{\partial \left( \mathcal{B}_k \right)^\alpha }{\partial P_\beta} \right] z dP_\alpha dP_\beta \\
    &\quad + \hbar \bar{z} (\mathcal{B}_x)_\alpha dX^\alpha \wedge dz + \hbar  (\mathcal{B}_x)_\alpha z dX^\alpha \wedge d\bar{z} \\
    &\quad + \hbar \bar{z} (\mathcal{B}_k)^\alpha dP_\alpha \wedge dz + \hbar  (\mathcal{B}_k)^\alpha z dP_\alpha \wedge d\bar{z} \\
    &\quad+ \mathcal{O}(\hbar^2).
\end{split}
\end{equation}
The effective ray equations in noncanonical coordinates $(X, P, z, \bar{z})$ are
\begin{subequations}
\begin{align} 
    \begin{split}
    \dot{X}^\mu &= \frac{\partial H}{\partial P_\mu} +\hbar \dot{X}^\nu \bar{z} \left[ \frac{\partial \left( \mathcal{B}_k \right)^\mu }{\partial X^\nu} - \frac{\partial \left( \mathcal{B}_x \right)_\nu }{\partial P_\mu} \right] z \\
    & \qquad+ \hbar \dot{P}_\nu \bar{z} \left[\frac{\partial \left( \mathcal{B}_k \right)^\mu }{\partial P_\nu} - \frac{\partial \left( \mathcal{B}_k \right)^\nu }{\partial P_\mu} \right] z \\
    & \qquad+ \hbar \bar{z} (\mathcal{B}_k)^\mu \dot{z} + \hbar \dot{\bar{z}} (\mathcal{B}_k)^\mu z 
    \end{split} \label{eq:dotX} \\
    \begin{split}
    \dot{P}_\mu &= -\frac{\partial H}{\partial X_\mu} + e \dot{X}^\nu F_{\nu \mu} \\
    &\qquad - \hbar \dot{X}^\nu \bar{z} \left[ \frac{\partial \left( \mathcal{B}_x \right)_\mu}{\partial X^\nu} - \frac{\partial \left( \mathcal{B}_x \right)_\nu}{\partial X^\mu} \right] z \\
    &\qquad - \hbar \dot{P}_\nu \bar{z} \left[  \frac{\partial \left( \mathcal{B}_x \right)_\mu }{\partial P_\nu} - \frac{\partial \left( \mathcal{B}_k \right)^\nu }{\partial X^\mu} \right] z \\
    &\qquad - \hbar \bar{z} (\mathcal{B}_x)_\mu \dot{z} - \hbar \dot{\bar{z}} (\mathcal{B}_x)_\mu z 
    \end{split} \label{eq:dotP}\\
    \dot z &= i \left[ \dot{X}^\alpha (\mathcal{B}_x)_\alpha + \dot{P}_\alpha (\mathcal{B}_k)^\alpha   - \frac{e}{2} F_{\mu \nu} s^{\mu \nu}  \right] z, \label{eq:zdot_01}\\
    \dot{\bar{z}} &= -i \bar{z} \left[ \dot{X}^\alpha (\mathcal{B}_x)_\alpha + \dot{P}_\alpha (\mathcal{B}_k)^\alpha   - \frac{e}{2} F_{\mu \nu} s^{\mu \nu}  \right].  \label{eq:zbdot_01}
\end{align}
\end{subequations}
Inserting the expressions of $\dot{z}$ and $\dot{\bar{z}}$ back into the equations for $\dot{X}^\mu$ and $\dot{P}_\mu$, we obtain
\begin{subequations}
\begin{align} 
    \begin{split}
    &\dot{X}^\mu = \frac{\partial H}{\partial P_\mu} +\hbar \dot{X}^\nu \bar{z} (\mathcal{F}_{k x})\indices{_\nu^\mu} z + \hbar \dot{P}_\nu \bar{z} (\mathcal{F}_{k k})^{\nu \mu} z \\
    &\qquad - \frac{i e \hbar}{2} \bar{z} [(\mathcal{B}_k)^\mu, F_{\alpha \beta} s^{\alpha \beta}] z
    \end{split}\\
    \begin{split}
    &\dot{P}_\mu = -\frac{\partial H}{\partial X_\mu} + e \dot{X}^\nu F_{\nu \mu}  - \hbar \dot{X}^\nu \bar{z} (\mathcal{F}_{x x})_{\nu \mu} z \\
    &\qquad - \hbar \dot{P}_\nu \bar{z} (\mathcal{F}_{x k})\indices{^\nu_\mu} z + \frac{i e \hbar}{2} \bar{z} [(\mathcal{B}_x)_\mu, F_{\alpha \beta} s^{\alpha \beta}] z. 
    \end{split}
\end{align}
\end{subequations}
Since the ray equations are correct up to error terms of $\mathcal{O}(\hbar^2)$, we can replace $\dot{X}^\mu = P^\mu + \mathcal{O}(\hbar^1)$ and $\dot{P}_\mu = \Gamma^\alpha_{\beta \mu} P_\alpha P^\beta + e P^\nu F_{\nu \mu} + \mathcal{O}(\hbar^1)$ on the right-hand side in the above equations. The same replacement can be made in the equations of $z$ and $\Bar{z}$, which shows that Eqs. \eqref{eq:zdot_01} and \eqref{eq:zbdot_01} are equivalent to Eq. \eqref{eq:transp_zdot}. Furthermore, using the expressions for the components of the Berry curvature given in Eq. \eqref{eq:Berry_comp}, we can simplify some terms:
\begin{subequations}
\begin{align}
    \begin{split}
    &\hbar \dot{X}^\nu \bar{z} (\mathcal{F}_{k x})\indices{_\nu^\mu} z + \hbar \dot{P}_\nu \bar{z} (\mathcal{F}_{k k})^{\nu \mu} z \\
    &= \frac{e}{m^2} P^\alpha F_{\alpha \nu} S^{\nu \mu} + \mathcal{O}(\hbar^2), 
    \end{split} \\
    \begin{split}
    &\hbar \dot{X}^\nu \bar{z} (\mathcal{F}_{x x})_{\nu \mu} z + \hbar \dot{P}_\nu \bar{z} (\mathcal{F}_{x k})\indices{^\nu_\mu} z \\
    &= \frac{1}{2} R_{\mu \nu \alpha \beta} P^\nu S^{\alpha \beta} + \frac{e}{m^2} P^\alpha P_\rho \Gamma^\rho_{\mu \beta} F_{\alpha \nu} S^{\beta \nu} + \mathcal{O}(\hbar^2). \end{split}
\end{align}
\end{subequations}
Thus, the effective ray equations in noncanonical coordinates can be written in the simplified form
\begin{subequations}
\begin{align} 
    \begin{split}
    \dot{X}^\mu &= P^\mu + \frac{e }{m^2} P^\alpha F_{\alpha \nu} S^{\nu \mu}  \\
    &\qquad+ \frac{e \hbar}{2} F_{\alpha \beta} \bar{z} \left( \vnabla{}^\mu s^{\alpha \beta} - i [(\mathcal{B}_k)^\mu, s^{\alpha \beta}] \right) z \label{eq:dotX_2}
    \end{split} \\
    \begin{split}
    \dot{P}_\mu &= \Gamma^\alpha_{\beta \mu} P_\alpha P^\beta + e \dot{X}^\nu F_{\nu \mu} - \frac{1}{2} R_{\mu \nu \alpha \beta} P^\nu S^{\alpha \beta} \\
    &\qquad + \frac{e}{m^2} P^\alpha P_\rho \Gamma^\rho_{\mu \beta} F_{\alpha \nu} S^{\beta \nu} - \frac{e}{2} F_{\alpha \beta, \mu} S^{\alpha \beta} \\
    &\qquad - \frac{e \hbar}{2} F_{\alpha \beta} \bar{z} \left( s\indices{^{\alpha \beta}_{,\mu}} -i [(\mathcal{B}_x)_\mu, s^{\alpha \beta}] \right) z , \label{eq:dotP_2}        
    \end{split} \\
    \dot z &= i \left[ \dot{X}^\alpha (\mathcal{B}_x)_\alpha + \dot{P}_\alpha (\mathcal{B}_k)^\alpha   - \frac{e}{2} F_{\mu \nu} s^{\mu \nu}  \right] z, \\
    \dot{\bar{z}} &= -i \bar{z} \left[ \dot{X}^\alpha (\mathcal{B}_x)_\alpha + \dot{P}_\alpha (\mathcal{B}_k)^\alpha   - \frac{e}{2} F_{\mu \nu} s^{\mu \nu}  \right].
\end{align}
\end{subequations}
In Eqs. \eqref{eq:dotX_2} and \eqref{eq:dotP_2}, the terms involving the derivatives of $ s^{\alpha \beta}$ can be rewritten as in Appendix \ref{app:noname}. Then we are left with the more compact form of the equations
\begin{subequations}
\begin{align} 
    &\dot{X}^\mu = P^\mu, \\
    \begin{split}
    &\dot{X}^\nu \nabla_\nu \dot{P}_\mu = e \dot{X}^\nu F_{\nu \mu} - \frac{1}{2} R_{\mu \nu \alpha \beta} P^\nu S^{\alpha \beta} \\ & \qquad \qquad\qquad- \frac{e}{2} S^{\alpha \beta} \nabla_\mu F_{\alpha \beta} , \end{split}\\
    &\dot z = i \left[ \dot{X}^\alpha (\mathcal{B}_x)_\alpha + \dot{P}_\alpha (\mathcal{B}_k)^\alpha   - \frac{e}{2} F_{\mu \nu} s^{\mu \nu}  \right] z, \label{eq:dotz}\\
    &\dot{\bar{z}} = -i \bar{z} \left[ \dot{X}^\alpha (\mathcal{B}_x)_\alpha + \dot{P}_\alpha (\mathcal{B}_k)^\alpha   - \frac{e}{2} F_{\mu \nu} s^{\mu \nu}  \right] . \label{eq:dotzb}
\end{align}
\end{subequations}
Note that the equation for $X^\mu$ is independent of the internal degrees of freedom $z$ and $\Bar{z}$, while the equation for $P_\mu$ depends on the internal degrees of freedom only through the spin tensor $S^{\alpha \beta}$. Thus, we can replace the evolution equations for the internal degrees of freedom $z$ and $\Bar{z}$ with an evolution equation for the spin tensor $S^{\alpha \beta}$. We start by expanding the covariant derivative of the spin tensor as
\begin{equation}
    \dot{X}^\nu \nabla_\nu S^{\alpha \beta} = \dot{S}^{\alpha \beta} + \dot{X}^\nu \left( \Gamma^\alpha_{\nu \gamma} S^{\gamma \beta} + \Gamma^\beta_{\nu \gamma} S^{\alpha \gamma} \right).
\end{equation}
Recall that $S^{\alpha \beta} = \hbar \Bar{z} s^{\alpha \beta} z$, where $s^{\alpha \beta}$ depends on the eigenspinors $\Sigma_A$ and therefore is a function of $X^\mu$ and $P_\mu$. Then, we have
\begin{equation}
    \dot{s}^{\alpha \beta} = \dot{X}^\nu \partial_\nu s^{\alpha \beta} + \dot{P}_\nu \vnabla{}^\nu s^{\alpha \beta},
\end{equation}
and the evolution equation for the spin tensor becomes
\begin{equation}
\begin{split}
    \dot{X}^\nu \nabla_\nu S^{\alpha \beta} &= \hbar \bigg[ \dot{\Bar{z}} s^{\alpha \beta} z + \Bar{z} s^{\alpha \beta} \Dot{z} + \dot{X}^\nu \Bar{z} (\nabla_\nu s^{\alpha \beta} ) z \\
    & \qquad \qquad + \dot{P}^\nu \Bar{z} (\vnabla{}^\nu s^{\alpha \beta} ) z \bigg].
\end{split}
\end{equation}
Next, using Eqs. \eqref{eq:dotz} and \eqref{eq:dotzb}, together with the relations derived in Appendix \ref{app:noname}, we arrive at
\begin{equation}
    \dot{X}^\nu \nabla_\nu S^{\alpha \beta} = \frac{i e}{8} F_{\mu \nu} \Bar{z}_A \Bar{\Sigma}_A [\sigma^{\mu \nu}, \sigma^{\alpha \beta}] \Sigma_B z_B.
\end{equation}
The commutator in the above equation can be calculated using the properties of the gamma matrices, and we obtain
\begin{equation}
    \frac{i}{2} [\sigma^{\mu \nu}, \sigma^{\alpha \beta}] = g^{\nu \alpha} \sigma^{\mu \beta} - g^{\mu \alpha} \sigma^{\nu \beta} - g^{\nu \beta} \sigma^{\mu \alpha} + g^{\mu \beta} \sigma^{\nu \alpha}.
\end{equation}
Finally, we can write the gauge-invariant gravitational spin Hall equations in terms of the variables $\{X^\mu, P_\mu, S^{\alpha \beta} \}$ as
\begin{subequations} \label{eq:GSHE_massive}
\begin{align} 
    \dot{X}^\mu &= P^\mu, \label{eq:dotX_massive} \\
    \begin{split}
    \dot{X}^\nu \nabla_\nu \dot{P}_\mu &= e \dot{X}^\nu F_{\nu \mu} - \frac{1}{2} R_{\mu \nu \alpha \beta} P^\nu S^{\alpha \beta} \label{eq:dotP_massive} \\
    & \qquad  - \frac{e}{2} S^{\alpha \beta} \nabla_\mu F_{\alpha \beta} , \end{split}\\
    \dot{X}^\nu \nabla_\nu S^{\alpha \beta} &= e F\indices{_\nu^\alpha} S^{\nu \beta} - e F\indices{_\nu^\beta} S^{\nu \alpha}. \label{eq:dotS_massive}
\end{align}
\end{subequations}
These equations describe the semiclassical motion of massive Dirac wave packets at linear order in spin. Compared to the lowest-order Lorentz force law derived in Eq. \eqref{eq:Lorentz_force1}, these equations contain additional spin-dependent terms that reflect the spin-orbit coupling between the external (represented by $X^\mu$ and $P_\mu$) and the internal (represented by $S^{\alpha \beta}$) degrees of freedom of the wave packet.

\subsection{Comparison with other results} \label{sec:MPD_comp}

In this section, we compare the gravitational spin Hall equations for massive Dirac wave packets derived above with other related results in the literature. We start by presenting a comparison with the Mathisson-Papapetrou equations for spinning bodies. For a charged compact object with conserved energy-momentum tensor, the Mathisson-Papapetrou equations are (ignoring quadrupole terms) \citep{Dixon74,Gralla2010}

\begin{subequations} \label{eq:MPD}
\begin{align}
    \begin{split}
    \dot{X}^\nu \nabla_\nu P_\mu &=  e \dot{X}^\nu F_{\nu \mu} - \frac{1}{2} R_{\mu \nu \alpha \beta} \dot{X}^\nu S^{\alpha \beta}  \\
    &\qquad- \frac{1}{2} Q^{\alpha \beta} \nabla_\mu F_{\alpha \beta}, 
    \end{split} \label{eq:MPD_P} \\
    \begin{split}
    \dot{X}^\nu \nabla_\nu S^{\alpha \beta} &= 2 P^{[\alpha} \dot{X}^{\beta]} + 2 Q^{\mu [\alpha} F\indices{^{\beta]}_\mu} ,
    \end{split} \label{eq:MPD_S}
\end{align}
\end{subequations}
where $Q^{\alpha \beta}$ is the body's electromagnetic dipole moment tensor. Comparison of Eq. \eqref{eq:dotP_massive} with Eq. \eqref{eq:MPD_P} shows that the dipole moment tensor has to be
\begin{equation}
    Q^{\alpha \beta} = e S^{\alpha \beta}.
\end{equation}
However, the Mathisson-Papapetrou equations are underdetermined and do not contain an evolution equation for the worldline $X^\mu$. This freedom can be fixed using a spin supplementary condition \citep{Costa2015}. In particular, the gravitational spin Hall equations \eqref{eq:GSHE_massive} are equivalent (up to linear order in spin) to the Mathisson-Papapetrou equations \eqref{eq:MPD} together with the Tulczyjew-Dixon spin supplementary condition
\begin{equation} \label{eq:S.p=0}
    S^{\alpha \beta} P_\beta = 0.
\end{equation}
In the context of the WKB analysis for the massive Dirac equation, this property is already satisfied by the spin tensor defined in Eq. \eqref{eq:Spin_tensor_massive}. The evolution equation for the worldline $X^\mu$ can be obtained by taking the covariant derivative $\dot{X}^\nu \nabla_\nu$ of the spin supplementary condition and using the Mathisson-Papapetrou equations. Ignoring terms that are quadratic in spin, we obtain
\begin{equation}
    \dot{X}^\mu = \frac{\dot{X}^\beta P_\beta}{P^\beta P_\beta} P^\mu.
\end{equation}
This equation is equivalent to Eq. \eqref{eq:dotX_massive} after we introduce a worldline parametrization for which 
\begin{equation}
    \dot{X}^\beta P_\beta = P^\beta P_\beta.
\end{equation}
It follows that $P^{[\alpha} \dot{X}^{\beta]} = 0$, and Eq. \eqref{eq:dotS_massive} is also equivalent to Eq. \eqref{eq:MPD_S}. Thus, when we ignore the quadrupole moments and terms quadratic in spin, the gravitational spin Hall equations \eqref{eq:GSHE_massive} are a particular case of the Mathisson-Papapetrou equations \eqref{eq:MPD} together with the Tulczyjew-Dixon spin supplementary condition $S^{\alpha \beta} P_\beta = 0$ and the electromagnetic dipole moment tensor $Q^{\alpha \beta} = e S^{\alpha \beta}$.

Similar conclusions regarding the equivalence of the Mathisson-Papapetrou equations and the semiclassical dynamics of massive Dirac fields have also been obtained in Refs. \citep{audretsch,rudiger,Cianfrani_2008,CIANFRANI2008}. However, these papers do not consider an external electromagnetic field $F_{\mu \nu}$.

\section{Massless Dirac fields} \label{sec:massless}

In this section, we present our semiclassical analysis of massless Dirac fields. In this case, the principal symbol is reduced to
\begin{equation}
    D|_{m=0} = D_0 = -\gamma^\mu k_\mu.
\end{equation}
The properties of the eigenspinors will result in a different Berry connection. We start in Sec. \ref{sec:WKB_0_massless} by analyzing the WKB equations at the lowest order. We derive the corresponding dispersion relation, and we introduce the eigenspinors of the principal symbol $D_0$. In Sec. \ref{sec:WKB_1_massless}, we derive the transport equation for the amplitude $\psi_0$, and we introduce the Berry connection and the Berry phase, which are closely related to the corresponding ones for electromagnetic and gravitational waves. Throughout this section, we will use the Weyl or chiral basis defined in Appendix \ref{app:massless_eigenspinors}.

\subsection{WKB equations at leading order} \label{sec:WKB_0_massless}

At the lowest order in $\hbar$ and setting $m = 0$, the Euler-Lagrange equations \eqref{eq:HJ_full} reduce to
\begin{subequations}
\begin{align}
    D_0 \psi_0 &= 0, \\
    \bar{\psi}_0 D_0 &= 0, \\
    \nabla_\mu {j_0}^\mu &= 0.
\end{align}
\end{subequations}
The first two equations admit nontrivial solutions $\psi_0$ if and only if the principal symbol matrix $D_0$ is singular. From this condition we obtain the dispersion relation
\begin{equation} \label{eq:dispersion_m=0}
    \det D_0 = 0 \qquad \Leftrightarrow \qquad k_\mu k^\mu = 0.
\end{equation}
Furthermore, $\psi_0$ needs to be an eigenspinor of $D_0$ with eigenvalue zero. Under the restriction given by the dispersion relation, together with the additional requirement that $k^\mu$ is future-directed with respect to the choice of orthonormal tetrad $(e_a)^\mu$, the principal symbol matrix $D_0$ has rank 2, and there are two eigenspinors with eigenvalue zero. Thus, we can write $\psi_0$ as
\begin{equation} \label{eq:psi0_m=0}
    \psi_0 (x, k) = \sqrt{\mathcal{I}_0(x)} z_A(x) \Sigma_A(x, k).
\end{equation}
The eigenspinors can be expressed in the Weyl basis as
\begin{equation}
    \Sigma_0 = \begin{pmatrix} 0 \\ u \end{pmatrix}, \qquad  \Sigma_1 = \begin{pmatrix} v \\ 0 \end{pmatrix},
\end{equation}
where $u$ and $v$ are $2$-spinors satisfying
\begin{equation} \label{eq:eigen_uv}
    k_\mu \sigma^\mu u = 0, \qquad k_\mu \bar{\sigma}^\mu v = 0.
\end{equation}
In the above equations, we use the spacetime Pauli $4$-vectors $\sigma^\mu = (e_a)^\mu \sigma^a$, $\bar{\sigma}^\mu = (e_a)^\mu \bar{\sigma}^a$, which are defined with the help of the tetrad $(e_a)^\mu$ and the flat spacetime Pauli $4$-vectors $\sigma^a = \left( \mathbb{I}_2, \sigma^i \right)$, $\bar{\sigma}^a = \left( \mathbb{I}_2, -\sigma^i \right)$.   
Since the eigenspinors satisfy
\begin{equation}
    \gamma^5 \Sigma_0 = \Sigma_0, \qquad \gamma^5 \Sigma_1 = -\Sigma_1,
\end{equation}
and we have chosen $k^\mu$ to be future-oriented, we can say that $\Sigma_0$ represents a right-hand chiral fermion of positive energy, while $\Sigma_1$ represents a left-hand chiral fermion of positive energy. 

Using the above relations, together with the properties of the eigenspinors listed in Appendix \ref{app:massless_eigenspinors}, the current ${j_0}^\mu $ can be expressed as
\begin{equation}
\begin{split}
    {j_0}^\mu &= \bar{\psi_0} \gamma^\mu \psi_0 \\
    &= \mathcal{I}_0 \left( \bar{z}_0 z_0 \bar{u} \sigma^\mu u + \bar{z}_1 z_1 \bar{v} \bar{\sigma}^\mu v \right) \\
    &= \mathcal{I}_0 \frac{k^\mu}{k_\alpha t^\alpha},
\end{split}
\end{equation}
where we defined a timelike vector as $t^\alpha = (e_0)^\alpha$. With this expression, we obtain the following transport equations for the intensity $\mathcal{I}_0$:
\begin{equation} \label{eq:transp0_m=0}
\nabla_\mu j_0^\mu = \nabla_\mu \left( \mathcal{I}_0 \frac{k^\mu}{k_\alpha t^\alpha} \right) = 0.
\end{equation}

Ray equations at the lowest order in $\hbar$ can be obtained exactly as in Sec. \ref{sec:ray0}, either by differentiating the dispersion relation \eqref{eq:dispersion_m=0},
\begin{equation}
    k^\mu \nabla_\mu k_\nu = e k^\mu F_{\mu \nu},
\end{equation}
or by solving the Hamilton-Jacobi equation for the phase function. The resulting ray equations are the same as in Sec. \ref{sec:ray0}, with the only difference that $k^\mu$ is now a null vector.

\subsection{WKB equations at next-to-leading order} \label{sec:WKB_1_massless}

For the massless case, the Euler-Lagrange equations \eqref{eq:HJ1_fullA} and \eqref{eq:HJ2_fullA} at order $\hbar^1$ only are
\begin{subequations}
\begin{align}
    D_0 \psi_1  &= - i  \gamma^\mu \nabla_\mu \psi_0, \\
    \Bar{\psi}_1 D_0  &=  i \nabla_\mu \Bar{\psi}_0 \gamma^\mu.
\end{align}
\end{subequations}
We treat this inhomogeneous system of linear algebraic equations exactly as in the massive case, and we obtain the solvability conditions
\begin{subequations}
\begin{align}
    \bar{\Sigma}_0 \gamma^\mu \nabla_\mu \psi_0 = \bar{\Sigma}_1 \gamma^\mu \nabla_\mu \psi_0 = 0, \\
    \nabla_\mu \Bar{\psi}_0 \gamma^\mu \Sigma_0 = \nabla_\mu \Bar{\psi}_0 \gamma^\mu \Sigma_1 = 0.
\end{align}
\end{subequations}
Using Eqs. \eqref{eq:psi0_m=0} and \eqref{eq:transp0_m=0}, the solvability conditions can be rewritten as a transport equation for $z$:
\begin{subequations}\label{eq:z_transp}
\begin{align} 
    k^\mu \nabla_\mu z_A &= i M_{A B} z_B, \\
    k^\mu \nabla_\mu \bar{z}_B &= -i \bar{z}_A M_{A B},
\end{align}   
\end{subequations}
where the $2 \times 2$ Hermitian matrix $M$ is defined as
\begin{equation}
    M = \frac{i k_\alpha t^\alpha}{2} \begin{pmatrix} \bar{u} \sigma^\mu \nabla_\mu u - \nabla_\mu \bar{u} \sigma^\mu u & 0 \\ 0 & \bar{v} \bar{\sigma}^\mu \nabla_\mu v - \nabla_\mu \bar{v} \bar{\sigma}^\mu v \end{pmatrix}.
\end{equation}
Note that, in contrast to the massive case, the matrix $M$ is diagonal. This reflects the fundamental difference from the Berry connection obtained in the massive. Using Eqs. \eqref{eq:2spinor_ortho}--\eqref{eq:2spinor_momentum}, we rewrite $M$ as
\begin{equation}
    M = k^\mu B_{\mu} - \frac{e}{2} F_{\mu \nu} s^{\mu \nu},
\end{equation}
where we introduced the Berry connection
\begin{equation} \label{eq:Berry_connection_massless}
    B_{\mu} = \frac{i}{2} \begin{pmatrix} \bar{u} \nabla_\mu u - \nabla_\mu \bar{u} u & 0 \\ 0 & \bar{v} \nabla_\mu v - \nabla_\mu \bar{v} v \end{pmatrix}
\end{equation}
and
\begin{equation}
    s^{\mu \nu} = \frac{i}{4} \left(\bar{u} \sigma^\mu v \bar{v} \sigma^\nu u - \bar{u} \sigma^\nu v \bar{v} \sigma^\mu u \right) \sigma_3,
\end{equation}
where $\sigma_3$ is the third Pauli matrix.

Since the matrix $M$ is diagonal, the dynamics of $z_0$ and $z_1$ is decoupled. The transport equation \eqref{eq:z_transp} can be integrated along a worldline $x^\mu(\tau)$, with $ \dot{x}^\mu = k^\mu$, and we obtain
\begin{equation}
    z_0(\tau) = e^{i \gamma_0(\tau)} z_0 (\tau_0), \qquad z_1(\tau) = e^{i \gamma_1(\tau)} z_1 (\tau_0),
\end{equation}
where
\begin{equation}
    \gamma_A(\tau) = \int_{\tau_0}^\tau d \tau' M_{A A}
\end{equation}
is the Berry phase. It follows that $\bar{z} z$ and $\bar{z} \sigma_3 z$ are conserved along integral curves of $k^\mu$. As in the case of electromagnetic \citep{GSHE2020} or gravitational waves \citep{GSHE_GW}, the Berry phase describes the dynamics of the internal spin degree of freedom and represents the next-to-leading order correction to the overall phase factor of the WKB ansatz.

The Berry connection in Eq. \eqref{eq:Berry_connection_massless} can also be related to the corresponding Berry connections for electromagnetic and gravitational waves obtained in Refs. \citep{GSHE2020,GSHE_GW}. First, notice that $s^{\mu \nu}$ is orthogonal to $k_\mu$ due to Eq. \eqref{eq:eigen_uv}, as well as to the timelike covector $t_\mu = (e_0)^\mu$. This last property follows from the definition of $\sigma^a$, with $\sigma^0 = \mathbb{I}_2$, and the orthogonality of the $2$-spinors given in Eq. \eqref{eq:2spinor_ortho}. Thus, it is convenient to define two complex null vectors orthogonal to $k_\mu$ and $t_\mu$:
\begin{equation}
    m^\mu = \frac{1}{\sqrt{2}} \bar{v} \sigma^\mu u, \qquad \bar{m}^\mu = \frac{1}{\sqrt{2}} \bar{u} \sigma^\mu v.
\end{equation}
Using the relations satisfied by the eigenspinors $u$ and $v$ given in Appendix \ref{app:massless_eigenspinors}, it is straightforward to show that $m^\mu m_\mu = \bar{m}^\mu \bar{m}_\mu = 0 $ and $m^\mu \bar{m}_\mu = 1$. Thus, the covectors $\{k_\alpha, t_\alpha, m_\alpha, \bar{m}_\alpha \}$ form a tetrad and we can write
\begin{equation}
    s^{\mu \nu} = i \bar{m}^{[\mu} m^{\nu]} \sigma_3.
\end{equation}
Note that the tensor $s^{\mu \nu}$ is independent of the actual choice of $m_\alpha$ and $\bar{m}_\alpha$, and is fully determined by $k_\alpha$ and $t_\alpha$:
\begin{equation}
    s^{\mu \nu} = - \frac{1}{2} \frac{ \varepsilon^{\alpha \beta \gamma \lambda} k_\gamma t_\lambda }{ \sqrt{k \cdot k + (k \cdot t)^2}} \sigma_3.
\end{equation}
In the present case, $k \cdot k = 0$ and the above relation can be simplified, but we give the most general expression since it will become important in the following section to consider the possibility of a non-null $k_\alpha$. 

Furthermore, using the relation between the eigenspinors $u$ and $v$ given in Eq. \eqref{eq:conj}, we can relate the diagonal components of the Berry connection in Eq. \eqref{eq:Berry_connection_massless} as
\begin{equation}
    \bar{u}\nabla_{\mu}u -\nabla_{\mu}\bar{u}u + i\nabla_{\mu}\theta = -(\bar{v}\nabla_{\mu}v -\nabla_{\mu}\bar{v}v + i\nabla_{\mu}\theta ),
\end{equation}
where $\theta$ is the phase factor introduced in Eq. \eqref{eq:conj}. Then, it follows that the Berry connection can be expressed in terms of the complex null vectors introduced above:
\begin{equation}
\begin{split}
    B_{\mu} &= \frac{i}{4}(\bar{m}^{\alpha} \nabla_{\mu} m_{\alpha} - m^{\alpha} \nabla_{\mu} \bar{m}_{\alpha})\sigma_{3} + \frac{1}{2}(\nabla_{\mu}\theta) \mathbb{I}_2 \\
    &= \frac{i}{2} \bar{m}^{\alpha} \nabla_{\mu} m_{\alpha} \sigma_{3} + \frac{1}{2}(\nabla_{\mu}\theta) \mathbb{I}_2
\end{split}
\end{equation}
The first term on the right-hand side has the same form as the Berry connection for electromagnetic waves \citep[Eq. 3.42]{GSHE2020}, except for a proportionality factor of $\frac{1}{2}$, which accounts for the fact that here we are dealing with a spin-$\frac{1}{2}$ field. Furthermore, remember that the eigenspinors $u$ and $v$, as well as the complex null vectors $m^\alpha$ and $\bar{m}^\alpha$, are functions of $x^\mu$ and $k_\mu$. Then, when applying the chain rule as in Eq. \eqref{eq:chain_rule}, we obtain
\begin{subequations}
\begin{align}
    k^\mu \nabla_\mu \left[ m_\alpha (x, k) \right] &= k^\mu \hnabla_\mu m_\alpha + e k^\mu F_{\mu \nu} \vnabla{}^\nu m_\alpha, \\
    k^\mu \nabla_\mu \left[ \theta (x, k) \right] &= k^\mu \hnabla_\mu \theta + e k^\mu F_{\mu \nu} \vnabla{}^\nu \theta,
\end{align}
\end{subequations}
and the Berry connection contracted with $k^\mu$ can be written as
\begin{equation}
\begin{split}
    k^\mu B_{\mu} &= \frac{i}{2} k^\mu \left( \bar{m}^{\alpha} \hnabla_{\mu} m_{\alpha}  + e F_{\mu \nu} \bar{m}^{\alpha} \vnabla{}^{\mu} m_{\alpha} \right) \sigma_{3} \\
    &\quad+ \frac{1}{2}k^\mu ( \hnabla_{\mu} \theta + e F_{\mu \nu} \vnabla{}^\nu \theta ) \mathbb{I}_2.
\end{split}
\end{equation}
Although the above form of the Berry connection is similar to the corresponding ones for electromagnetic \citep[Eq. 3.42]{GSHE2020} and gravitational \citep[Eq. 3.32]{GSHE_GW} waves, there are two additional terms. The first additional term contains a vertical derivative of $m_\alpha$ and is present because here we consider charged Dirac fields in an external electromagnetic $F_{\mu \nu}$. The second additional term simply encodes an additional phase degree of freedom $\theta$ when choosing the eigenspinors $u$ and $v$, in comparison to the complex null vectors $m^\mu$ and $\bar{m}^\mu$ that define the circular polarization basis in the case of electromagnetic and gravitational waves. However, in the following we will see that this additional phase does not contribute to the Berry curvature or to the equations of motion.

\subsection{Geometric definition of the Berry connection and Berry curvature} \label{sec:geometry_Berry0}

Similar to the massive case, we can redefine the Berry connection as a connection on the Lagrangian submanifold. The Hamiltonian vector field $X_H$ is the same as in Eq. \eqref{eq:X_H}, and we can use
\begin{equation}
    \mathcal{B}(X_H) = k^\mu B_{\mu}
\end{equation}
to obtain the Berry connection $\mathcal{B}$ defined on the Lagrangian submanifold as
\begin{equation}
\begin{split}
    \mathcal{B} &= \left[ \frac{i}{2} \bar{m}^{\alpha} \nabla_{\mu} m_{\alpha} \sigma_{3} + \frac{1}{2} (\nabla_\mu \theta) \mathbb{I}_2 \right] d x^\mu \\
    &\qquad+ \left[ \frac{i}{2} \bar{m}^{\alpha} \vnabla{}^{\mu} m_{\alpha} \sigma_{3} + \frac{1}{2} (\vnabla{}^\mu \theta) \mathbb{I}_2 \right] d k_\mu.
\end{split}
\end{equation}
This is a Lie algebra-valued one-form defined on the Lagrangian submanifold, where the corresponding Lie algebra is $\mathfrak{u}(1) \times \mathfrak{u}(1)$. The Berry curvature of this connection can be calculated using Eq. \eqref{eq:def_curvature}, and we obtain
\begin{equation}
\begin{split}
    \mathcal{F} &= (\mathcal{F}_{x x})_{\mu \nu} dx^\mu dx^\nu + (\mathcal{F}_{k x})\indices{_\mu^\nu} dx^\mu dk_\nu \\
    &\qquad + (\mathcal{F}_{x k})\indices{^\mu_\nu} dk_\mu dx^\nu + (\mathcal{F}_{k k})_{\mu \nu} dk^\mu dk^\nu, 
\end{split}
\end{equation}
where
\begin{subequations}
\begin{align}
    \left( \mathcal{F}_{x x} \right)_{\mu \nu} &= \frac{\partial \left( \mathcal{B}_x \right)_\nu}{\partial x^\mu} - \frac{\partial \left( \mathcal{B}_x \right)_\mu}{\partial x^\nu} , \\
    \left( \mathcal{F}_{k k} \right)^{\mu \nu} &= \frac{\partial \left( \mathcal{B}_k \right)^\nu }{\partial k_\mu} - \frac{\partial \left( \mathcal{B}_k \right)^\mu }{\partial k_\nu} , \\
    \left( \mathcal{F}_{k x} \right)\indices{_\mu^\nu} = -\left( \mathcal{F}_{x k} \right)\indices{^\nu_\mu} &= \frac{\partial \left( \mathcal{B}_k \right)^\nu }{\partial x^\mu} - \frac{\partial \left( \mathcal{B}_x \right)_\mu }{\partial k_\nu} ,
\end{align}
\end{subequations}
and
\begin{subequations}
\begin{align}
    \left( \mathcal{B}_{x} \right)_{\mu} &= \frac{i}{2} \bar{m}^{\alpha} \nabla_{\mu} m_{\alpha} \sigma_{3} + \frac{1}{2} (\nabla_\mu \theta) \mathbb{I}_2, \\
    \left( \mathcal{B}_{k} \right)\indices{^\mu} &= \frac{i}{2} \bar{m}^{\alpha} \vnabla{}^{\mu} m_{\alpha} \sigma_{3} + \frac{1}{2} (\vnabla{}^\mu \theta) \mathbb{I}_2 .
\end{align}
\end{subequations}
Note that terms that involve the phase function $\theta$ will not contribute to the Berry curvature. Thus, the Berry curvature will have a form similar to that of electromagnetic waves \citep{GSHE2020,Harte_2022}:
\begin{subequations} \label{eq:Berry_comp0}
\begin{align}
    \begin{split}
    \left( \mathcal{F}_{x x} \right)_{\mu \nu} &= \frac{s^{\alpha\beta}}{2}  \bigg[ - R_{\mu \nu \alpha  \beta} + \frac{2}{ k \cdot k + (k \cdot t)^2}  k_\rho \Gamma^\rho_{\alpha [ \nu }   \\ 
    &\qquad  \times\bigg( \Gamma^\sigma_{\mu] \beta} k_\sigma - 2 (k \cdot t)  \nabla_{\mu]} t_\beta  \bigg) \\
    &\qquad- \frac{2 k \cdot k}{ k \cdot k + (k \cdot t)^2} \nabla_{\mu} t_\alpha \nabla_{\nu} t_\beta\bigg], 
    \end{split} \\
    \left( \mathcal{F}_{k k} \right)^{\mu \nu} &= -\frac{s^{\mu \nu}}{k \cdot k + (k \cdot t)^2} , \\
    \left( \mathcal{F}_{k x} \right)\indices{_\mu^\nu} &= - \left( \mathcal{F}_{x k} \right)\indices{^\nu_\mu} \nonumber\\
    &= \frac{ s^{\gamma \nu}  }{k \cdot k + (k \cdot t)^2}  \left( k_\rho \Gamma^\rho_{\mu \gamma} -( k \cdot t ) \nabla_\mu t_\gamma \right).
\end{align}
\end{subequations}
However, there are some minor differences between the above expressions and the Berry curvature for electromagnetic waves used in Refs. \citep{GSHE2020,Harte_2022}. First of all, since here we are working with a spin-$\frac{1}{2}$ Dirac field, there is an additional factor of $\frac{1}{2}$ in the definition of $s^{\alpha \beta}$. Second, we allow the possibility that $k_\mu$ is not null, which leads to some additional terms. In fact, the calculation of the Berry connection in Ref. \citep[Appendix C]{GSHE2020} was done without assuming $k_\mu$ to be null and only in the final expressions was $k \cdot k = 0$ used. In the next section, it will become clear why we should not restrict the discussion to a null $k_\mu$.

\subsection{Effective dispersion relation and spin-orbit coupling}

In this section, we aim to derive the effective ray equations describing the gravitational spin Hall effect of massless Dirac wave packets. This can be achieved by taking into account the spin-orbit coupling between the external and the internal degrees of freedom. Given the similarities between the Berry connection derived above and the corresponding Berry connection for electromagnetic waves, we will account for spin-orbit couplings by following similar steps as in Refs. \citep[Sec. IV]{GSHE2020} and \citep[Sec. II]{Harte_2022}. The main idea behind this approach is to observe that, because of the diagonal form of the Berry connection, the dynamics of the massless Dirac spinors of right-handed and left-handed chirality is decoupled, and the WKB fields can be of the form
\begin{equation} \label{eq:init_spinor}
    \Psi = \sqrt{\mathcal{I}_0} \begin{pmatrix} 0 \\ u \end{pmatrix} e^{i \gamma_0} e^{i S / \hbar} \quad \text{or} \quad \Psi = \sqrt{\mathcal{I}_0} \begin{pmatrix} v \\ 0 \end{pmatrix} e^{i \gamma_1} e^{i S / \hbar}.
\end{equation}
The above fields have a total phase function $\tilde{S} = S + \hbar \gamma_A$, where the Berry phase $\gamma_A$ represents a higher-order correction to the leading-order phase function $S$. Spin-orbit couplings can be accounted for by treating $S$ and $\gamma_A$ on equal footing and using the total phase function $\tilde{S}$ to define an effective dispersion relation. Using the results obtained in the previous sections, we can write
\begin{equation}
\begin{split}
    &\frac{1}{2} (\nabla_\mu \tilde{S} + e A_\mu) (\nabla^\mu \tilde{S} + e A^\mu) \\
    &\qquad- \hbar (\nabla^\mu S + e A^\mu) (\nabla_\mu \gamma_A) = \mathcal{O}(\hbar^2).
\end{split}
\end{equation}
Using the definition of the Berry phase and introducing the notation $\tilde{k}_\mu = \nabla_\mu \tilde{S}+ e A_\mu$, we can rewrite the above equation as
\begin{equation}
    \frac{1}{2} \tilde{k}_\mu \tilde{k}^\mu - \hbar \bar{z} \left( \tilde{k}^\mu B_\mu - \frac{e}{2} F_{\mu \nu} s^{\mu \nu} \right) z = \mathcal{O}(\hbar^2),
\end{equation}
where $z = (e^{i \gamma_0} \quad 0)^\mathrm{T}$ or $ (0 \quad e^{i \gamma_1})^\mathrm{T}$, depending on the initial chirality of the spinor field $\Psi$ in Eq. \eqref{eq:init_spinor}. This is an effective dispersion relation containing spin-dependent correction terms to the leading-order dispersion relation obtained in Eq. \eqref{eq:dispersion_m=0}. Ray equations can be obtained by treating the effective dispersion relation as a Hamilton-Jacobi equation for the total phase function $\tilde{S}$. Using the method of characteristics, the Hamilton-Jacobi equation can be solved in terms of the ray equations determined by the Hamiltonian function
\begin{equation}
\begin{split}
    H(x, p) &= \frac{1}{2} (p_\mu + e A_\mu) (p^\mu + e A^\mu) \\
    &\quad- \hbar \bar{z} \left[ (p^\mu + e A^\mu) B_\mu - \frac{e}{2} F_{\mu \nu} s^{\mu \nu} \right] z.
\end{split}
\end{equation}
Assuming the symplectic two-form 
\begin{equation}
    \Omega = dx^\alpha \wedge dp_\alpha,
\end{equation}
we can derive the corresponding Hamilton's equations
\begin{equation}
    \dot{x}^\mu = \frac{\partial H}{\partial p_\mu}, \qquad \dot{p}_\mu = - \frac{\partial H}{\partial x^\mu}.
\end{equation}
These equations contain spin-dependent terms and can be viewed as a description of the gravitational spin Hall effect. Note that, compared to the effective ray equations \eqref{eq:ray_eq_1} or \eqref{eq:GHSE_massive_2} in the massive case, here there is no evolution equation for the internal degree of freedom $z$. This is because of the fundamental differences of the Berry connections in the massive and massless case. Here, the dynamics of $z$ is trivial and, up to a phase function $\gamma_A$, fixed by the initial conditions. However, both the Hamiltonian and the equations of motion contain gauge-dependent terms: on the one hand, due to the presence of the electromagnetic vector potential $A_\mu$, but also due to the Berry connection $B_\mu$, which depends on the choice of eigenspinors $u$ and $v$. We follow similar steps as in the massive case, and we remove the gauge-dependent terms by a series of coordinate transformations.

We start with a first coordinate transformation $(x, p) \mapsto (x, k)$, with $k_\mu = p_\mu + e A_\mu$. This has the effect of eliminating the electromagnetic vector potential $A_\mu$ from the Hamiltonian, and we obtain
\begin{equation}
    H(x, k) = \frac{1}{2} k_\mu k^\mu - \hbar \bar{z} \left[ k^\mu B_\mu - \frac{e}{2} F_{\mu \nu} s^{\mu \nu} \right] z.
\end{equation}
The symplectic two-forms in the new coordinates $(x, k)$ is
\begin{equation}
    \Omega = e F_{\alpha \beta} dx^\alpha dx^\beta + dx^\alpha \wedge dk_\alpha,
\end{equation}
and we can write Hamilton's equations as
\begin{equation}
    \dot{x}^\mu = \frac{\partial H}{\partial k_\mu}, \qquad  \dot{k}_\mu = -\frac{\partial H}{\partial x^\mu} + e \dot{x}^\nu F_{\nu \mu}.
\end{equation}
We have successfully eliminated the gauge-dependent vector potential $A_\mu$ from the Hamiltonian, and now the effect of the external electromagnetic field is described in a gauge-invariant way by the presence of the field strength tensor $F_{\alpha \beta}$ in the symplectic two-form $\Omega$.

Gauge-dependent terms related to the Berry connection $B_\mu$ can be eliminated by performing a second coordinate transformation $(x, k) \mapsto (X, K)$. This type of transformation was first introduced in Ref. \citep{Littlejohn1991}, and in the present case we define it as
\begin{subequations}
\begin{align}
    X^\mu &= x^\mu + \hbar \bar{z} (\mathcal{B}_k)^\mu z, \\
    K_\mu &= k_\mu - \hbar \bar{z} (\mathcal{B}_x)_\mu z - e \hbar F_{\mu \nu} \bar{z} (\mathcal{B}_k)^\nu z .
\end{align}
\end{subequations}
Following the same steps as in the massive case, the Hamiltonian in the new coordinates $(X, K)$ becomes
\begin{equation}
    H(X, K) = \frac{1}{2}g^{\alpha \beta} K_\alpha K_\beta + \frac{e}{2} F_{\alpha \beta} S^{\alpha \beta} + \mathcal{O}(\hbar^2).
\end{equation} 
Note that the effective dispersion relation $H = 0$ implies that, up to error terms of order $\hbar^2$, $K_\alpha$ is generally not null. This is why we avoided assuming null vectors in the previous section. In the above equation, we have introduced the spin tensor
\begin{equation} \label{eq:spin_tensor_massless}
    S^{\alpha \beta} = \hbar \bar{z} s^{\alpha \beta} z = - \hbar s \frac{ \varepsilon^{\alpha \beta \gamma \lambda} K_\gamma t_\lambda }{\sqrt{ K \cdot K + (K \cdot t)^2}},
\end{equation}
where $s = \frac{1}{2} \bar{z} \sigma^3 z = \pm \frac{1}{2}$ depending on the initial chirality of the Dirac fields. The spin tensor encodes the intrinsic angular momentum carried by the wave packet. The symplectic two-form in the $(X, K)$ coordinates can be written as
\begin{equation}
\begin{split}
    &\Omega = e F_{\alpha \beta} dX^\alpha dX^\beta + dX^\alpha \wedge dK_\alpha \\
    &\quad - \hbar \bar{z} \left( \mathcal{F}_{x x} \right)_{\alpha \beta} z dX^\alpha dX^\beta - \hbar \bar{z} \left( \mathcal{F}_{k k} \right)^{\alpha \beta} z dK_\alpha dK_\beta \\
    &\quad - \hbar \bar{z}  \left( \mathcal{F}_{k x} \right)\indices{_\alpha^\beta} z \left( dX^\alpha dK_\beta - dK_\beta dX^\alpha \right) + \mathcal{O}(\hbar^2).
\end{split}
\end{equation}
The above equation can be written more compactly as
\begin{equation}
    \Omega =  dX^\alpha \wedge dK_\alpha + e F - \hbar \bar{z} \mathcal{F} z.
\end{equation}
Thus, we have arrived at a gauge-invariant description, where the Hamiltonian does not depend on the gauge fields $A_\mu$ and $B_\mu$, and the symplectic two-form now includes contributions from the electromagnetic field strength tensor $F_{\alpha \beta}$ and from the Berry curvature $\mathcal{F}$. The gauge-invariant equations of motion describing the gravitational spin Hall effect can now be written as
\begin{subequations}
\begin{align} 
    \begin{split}
    \dot{X}^\mu &= \frac{\partial H}{\partial K_\mu} -\hbar \dot{X}^\nu \bar{z} \left( \mathcal{F}_{x k} \right)\indices{^\mu_\nu} z \\
    &\qquad- \hbar \dot{K}_\nu \bar{z} \left( \mathcal{F}_{k k} \right)^{\mu \nu} z 
    \end{split} \\
    \begin{split}
    \dot{K}_\mu &= -\frac{\partial H}{\partial X^\mu} + e \dot{X}^\nu F_{\nu \mu} + \hbar \dot{X}^\nu \bar{z} \left( \mathcal{F}_{x x} \right)_{\mu \nu} z \\
    &\qquad + \hbar \dot{K}_\nu \bar{z} \left( \mathcal{F}_{k x} \right)\indices{_\mu^\nu} z. 
    \end{split}
\end{align}
\end{subequations}
After taking the derivatives of the Hamiltonian and inserting the expressions for the Berry curvature, we obtain
\begin{subequations}
\begin{align} 
    \begin{split}
    &\dot{X}^\mu = K^\mu + \frac{K \cdot t}{K \cdot K + (K \cdot t)^2} S^{\mu \alpha} K^\nu \nabla_\nu t_\alpha \\
    &- \frac{e}{K \cdot K + (K \cdot t)^2} \bigg[ \hbar s \sqrt{K \cdot K + (K \cdot t)^2} (\star F)^{\mu \sigma} t_\sigma \\
    & +\frac{1}{2} F_{\alpha \beta} S^{\alpha \beta} \left(K^\mu +  (K \cdot t)  t^\mu \right) - K^\sigma F_{\sigma \nu} S^{\mu \nu} \bigg],
    \end{split}\\
    \begin{split}
    &\dot{X}^\nu \nabla_\nu K_\mu = e \dot{X}^\nu F_{\nu \mu} - \frac{1}{2} R_{\mu \nu \alpha \beta} K^\nu S^{\alpha \beta} - \frac{e}{2} S^{\alpha \beta} \nabla_\mu F_{\alpha \beta} \\
     &+ \frac{e \nabla_\mu t_\sigma}{K \cdot K + (K \cdot t)^2} \bigg[ \hbar s \sqrt{K \cdot K + (K \cdot t)^2} (\star F)^{\sigma \nu} K_\nu \\
     & + \frac{K \cdot t}{2} F_{\alpha \beta} S^{\alpha \beta} K^\sigma  - (K \cdot t) K^\delta F_{\delta \nu} S^{\sigma \nu} \bigg].
     \end{split}
\end{align}
\end{subequations}
Furthermore, we can simplify the terms in square brackets. The first term can be rewritten as
\begin{align}
    \begin{split}
    &\frac{e}{K \cdot K + (K \cdot t)^2} \bigg[ \hbar s \sqrt{K \cdot K + (K \cdot t)^2} (\star F)^{\mu \sigma} t_\sigma \\
    & \qquad+\frac{1}{2} F_{\alpha \beta} S^{\alpha \beta} \left(K^\mu +  (K \cdot t)  t^\mu \right) - K^\sigma F_{\sigma \nu} S^{\mu \nu} \bigg] \\
    &= -\frac{e K \cdot t }{K \cdot K + (K \cdot t)^2} t^\sigma F_{\sigma \nu} S^{\nu \mu},
    \end{split}
\end{align}
and the second term can be expressed as
\begin{align}
    \begin{split}
     &\frac{e \nabla_\mu t_\sigma}{K \cdot K + (K \cdot t)^2} \bigg[ \hbar s \sqrt{K \cdot K + (K \cdot t)^2} (\star F)^{\sigma \nu} K_\nu \\
     &\qquad + \frac{K \cdot t}{2} F_{\alpha \beta} S^{\alpha \beta} K^\sigma  - (K \cdot t) K^\delta F_{\delta \nu} S^{\sigma \nu} \bigg]  \\
     &= - \frac{e (K \cdot K) \nabla_\mu t_\sigma}{K \cdot K + (K \cdot t)^2} \left( \frac{1}{2} F_{\alpha \beta} S^{\alpha \beta} t^\sigma + t^\delta F_{\delta \nu} S^{\nu \sigma} \right) \\
     &= \mathcal{O}(\hbar^2)
     \end{split}
\end{align}
The above term is of order $\hbar^2$ since $K \cdot K = \mathcal{O}(\hbar)$ and $S^{\alpha \beta} = \mathcal{O}(\hbar)$, and we can ignore it. Finally, we can write the gravitational spin Hall equations in the compact form 
\begin{subequations} \label{eq:GSHE_massless}
\begin{align} 
    &\dot{X}^\mu = K^\mu + \frac{K \cdot t}{K \cdot K + (K \cdot t)^2} S^{\mu \alpha} \left( K^\nu \nabla_\nu t_\alpha + e F_{\alpha \sigma} t^\sigma \right), \label{eq:GSHE_massless_x} \\
    &\dot{X}^\nu \nabla_\nu K_\mu = e \dot{X}^\nu F_{\nu \mu} - \frac{1}{2} R_{\mu \nu \alpha \beta} K^\nu S^{\alpha \beta} - \frac{e}{2} S^{\alpha \beta} \nabla_\mu F_{\alpha \beta}.  \label{eq:GSHE_massless_p}
\end{align}
\end{subequations}
These equations are gauge-invariant and contain spin-dependent correction terms to the geodesic equations. In contrast to the massive case discussed in Sec. \ref{sec:massive}, these equation explicitly depend on the choice of a timelike vector field $t^\alpha$. This dependence is also encountered for other massless fields \citep{GSHE2020,GSHE_GW} and the timelike vector field $t^\alpha$ has the role of fixing the energy centroid of the wave packet \citep{Harte_2022}. Furthermore, while for the massive case we had an additional evolution equation for the spin tensor, here the spin tensor is already fixed (up to the choice of sign in $s = \pm \frac{1}{2}$) by Eq. \eqref{eq:spin_tensor_massless}.

\subsection{Comparison with other results} \label{sec:MPD_comp_massless}

In this section, we compare the gravitational spin Hall equations \eqref{eq:GSHE_massless} for massless Dirac wave packets with other known results in the literature. We start with a brief comparison with the equations of motion for the gravitational spin Hall effect of electromagnetic \citep{GSHE2020} and gravitational waves \citep{GSHE_GW}. Then, we show that Eq. \eqref{eq:GSHE_massless} can be viewed as a particular case of the Mathisson-Papapetrou equations, together with an appropriate choice of the supplementary spin condition.

There are two main differences between the massless Dirac wave packets considered here and the electromagnetic and gravitational wave packets discussed in Refs. \citep{GSHE2020,GSHE_GW,Harte_2022}. First, these are fields of different spins. The Dirac fields are spin-$\frac{1}{2}$, while electromagnetic and gravitational fields have spin-$1$ and spin-$2$, respectively. From the point of view of the gravitational spin Hall equations, this difference is encoded in the absolute value of the constant $s$ entering the definition of the spin tensor $S^{\alpha \beta}$ in Eq. \eqref{eq:spin_tensor_massless}. Second and more importantly, the Dirac field has an electric charge $e$ and the Dirac equation is considered here in a fixed electromagnetic field $F_{\alpha \beta}$. This leads to additional terms in the gravitational spin Hall equations. If we set $F_{\alpha \beta} = 0$, Eq. \eqref{eq:GSHE_massless} reduces to 
\begin{subequations} \label{eq:gshe_massless_f=0}
\begin{align} 
    \dot{X}^\mu &= K^\mu + \frac{1}{K \cdot t} S^{\mu \alpha} K^\nu \nabla_\nu t_\alpha , \\
    \dot{X}^\nu \nabla_\nu K_\mu &=  - \frac{1}{2} R_{\mu \nu \alpha \beta} K^\nu S^{\alpha \beta}.
\end{align}
\end{subequations}
These equations have the same form as those for electromagnetic and gravitational wave packets \citep[Eq. 2.17]{Harte_2022}, and most of the results from Ref. \citep{Harte_2022} also apply to this case. In particular, the above equations are a special case of the Mathisson-Papapetrou equations, together with the Corinaldesi-Papapetrou \citep{CP_ssc,Costa2015} spin supplementary condition $S^{\alpha \beta} t_\beta = 0$, a particular choice for the parametrization of the worldline and additional initial conditions. 

For $F_{\alpha \beta} \neq 0$, it can still be shown that Eq. \eqref{eq:GSHE_massless} are related to the Mathisson-Papapetrou equations \eqref{eq:MPD}. Equations \eqref{eq:MPD_P} and \eqref{eq:GSHE_massless_p} are equivalent after imposing $Q^{\alpha \beta} = e S^{\alpha \beta}$. An evolution equation for the worldline can be obtained from the Mathisson-Papapetrou equations by using a spin supplementary condition. In this case, we pick
\begin{equation} \label{eq:ssc_massless}
    S^{\alpha \beta} \left [ K_\beta + (K \cdot t) t_\beta \right] = 0.
\end{equation}
Taking the covariant derivative $\dot{X}^\nu \nabla_\nu$ of the spin supplementary condition, using the Mathisson-Papapetrou equation \eqref{eq:MPD_S} for the spin tensor and ignoring terms quadratic in spin, we obtain
\begin{equation}
\begin{split}
    \dot{X}^\mu &= \frac{ \dot{X}^\beta \left [ K_\beta + (K \cdot t) t_\beta \right] }{ K^\beta \left [ K_\beta + (K \cdot t) t_\beta \right] } K^\mu \\
    &\qquad+ \frac{K \cdot t}{K \cdot K + (K \cdot t)^2} S^{\mu \alpha} \left( K^\nu \nabla_\nu t_\alpha + e F_{\alpha \sigma} t^\sigma \right).
\end{split}
\end{equation}
This is equivalent to Eq. \eqref{eq:GSHE_massless_x} after we chose a parametrization of the worldline such that
\begin{equation}
    \dot{X}^\beta \left [ K_\beta + (K \cdot t) t_\beta \right] = K^\beta \left [ K_\beta + (K \cdot t) t_\beta \right].
\end{equation}
Furthermore, a direct calculation shows that the Mathisson-Papapetrou equation \eqref{eq:MPD_S} is satisfied by the spin tensor given in Eq. \eqref{eq:spin_tensor_massless}. Thus, the gravitational spin Hall equations \eqref{eq:GSHE_massless} can be viewed as a particular case of the Mathisson-Papapetrou equations \eqref{eq:MPD}.

It should be noted here that the gravitational spin Hall equations \eqref{eq:GSHE_massless}, as well as their relation to the Mathisson-Papapetrou equations, break down when $K_\alpha + (K \cdot t) t_\alpha$ is null. In this case, we have
\begin{equation}
    K \cdot K + (K \cdot t)^2 = 0,
\end{equation}
and second term in Eq. \eqref{eq:GSHE_massless_x} blows up. However, this happens only when there is a fine balance between the external electromagnetic field $F_{\alpha \beta}$ and the choice of a timelike vector field $t^\alpha$. Such blowups can always be avoided by defining the centroid of the wave packet with respect to a different timelike vector field $\tilde{t}^\alpha$. 

More intuition about the gravitational spin Hall effect can be gained by looking at numerical examples of spin-dependent trajectories in various spacetimes. For the $F_{\alpha \beta} = 0$ case, numerical examples of spin Hall trajectories in Schwarzschild and Kerr spacetimes can be found in Refs. \citep{GSHE2020,Oanceathesis,gshe_lensing}. Furthermore, the \textit{Mathematica} notebook given in Ref. \citep[Appendix A.7]{Oanceathesis} can be used to obtain spin Hall trajectories in arbitrary spacetimes. However, that only works for $F_{\alpha \beta} = 0$, and the notebook would need to be extended to take into account a nonzero electromagnetic field $F_{\alpha \beta}$.

\section{Conclusions}
\label{sec:conclusions}

We presented a semiclassical analysis for massive and massless Dirac fields on arbitrary background spacetimes and in the presence of a fixed electromagnetic field. Our approach is based on a WKB approximation, and the resulting equations have been investigated at the leading and next-to-leading order in the expansion parameter $\hbar$. The semiclassical dynamics is expressed in terms of a Berry connection, which governs the dynamics of the spin internal degree of freedom of the wave packet and the associated Berry curvature, which determines corrections to the motion of the wave packet. This results in a gravitational spin Hall effect, meaning that wave packets will generally follow spin-dependent trajectories when propagating in inhomogeneous gravitational and electromagnetic fields.

In the massive case, we have shown that the gravitational spin Hall equations \eqref{eq:GSHE_massive} are a particular case of the Mathisson-Papapetrou equations, together with the Tulczyjew-Dixon spin supplementary condition \eqref{eq:S.p=0}. In the absence of an external electromagnetic field, our results are in agreement with those obtained in Refs. \citep{audretsch,rudiger}, where similar methods have been used.

For massless Dirac wave packets, we have also shown that the gravitational spin Hall equations \eqref{eq:GSHE_massless} are a particular case of the Mathisson-Papapetrou equations, but this time with a different spin supplementary condition given in Eq. \eqref{eq:ssc_massless}. Furthermore, in the absence of an external electromagnetic field, the gravitational spin Hall equations take the same form and share the same properties as in the case of electromagnetic \citep{GSHE2020,Harte_2022} or gravitational waves \citep{GSHE_GW}. 

We believe that the results derived in this paper can be used in several applications. First, the semiclassical equations derived for massive Dirac fields could be used to study the dynamics of electrons and protons. Gravitational effects are expected to play an important role in scenarios that arise in relativistic quantum information \citep{alsing2009spin,PhysRevA.69.032113,Palmer2012}, as well as in experiments involving particle accelerators \citep{Laszlo2018,laszlo2019}. Second, chiral kinetic theory \citep{PhysRevLett.109.181602,stephanov2012chiral,HIDAKA2022103989} is emerging as a semiclassical method for studying many body systems in high-energy physics \citep{KHARZEEV20161}, condensed matter physics \citep{doi:10.1142/11475,PhysRevLett.109.181602,Burkov_2015,PhysRevLett.118.127601,Duval_chiral_fermions,DUVAL2015} and astrophysics \citep{liu2018chiral,KAMADA2022104016,PhysRevD.105.096019,Tomoya2021}. Since chiral kinetic theory is based on the semiclassical dynamics of point particles with spin, we expect that our results can also be used in this context. In particular, the Berry curvature plays a central role in chiral kinetic theory, and, thus, we expect that our covariant formulation of the Berry curvature could be of use in studies of chiral kinetic theory in curved spacetime.

\section*{Acknowledgments}

The authors are grateful to Lars Andersson, Ilya Dodin, Vatsal Dwivedi, Abraham Harte, J\'er\'emie Joudioux, Banibrata Mukhopadhyay, Daniel Ruiz, and Michael Stone for helpful discussions.

\appendix
\renewcommand{\theequation}{\thesection.\arabic{equation}}

\section{Horizontal and vertical derivatives} \label{app:derivative_TM}

Let $(x^\mu, p_\mu)$ be canonical coordinates on the cotangent bundle $T^*M$ and considering a spinor field $\Psi(x,p)$ defined on $T^*M$. The horizontal and vertical derivatives of $\Psi(x,p)$ can be defined by extending the definition presented in Ref. \citep[section 3.5]{sharafutdinov2012integral} for the horizontal and vertical derivatives of tensor fields:
\begin{subequations}
\begin{align}
\vnabla{}^\mu \Psi ={}& \frac{\partial}{\partial p_\mu} \Psi, \\
\hnabla_\mu \Psi ={}&  
\frac{\partial}{\partial x^\mu} \Psi - \frac{i}{4} \omega\indices{_\mu^{a b}} \sigma_{a b} \Psi + \Gamma^\sigma_{\mu \rho} p_\sigma \frac{\partial}{\partial p_\rho} \Psi.
\end{align}
\end{subequations}
Note that, in contrast to \citep[section 3.5]{sharafutdinov2012integral}, we have the opposite sign for the last term in the definition of the horizontal derivative. This is because we are considering fields defined on $T^*M$, and not on $TM$, as is the case in the reference mentioned earlier. The horizontal and vertical derivatives satisfy the following properties:
\begin{subequations}
\begin{align}
    [\hnabla{}_\mu, \vnabla{}^\nu] =0, \quad [\vnabla{}^\mu, \vnabla{}^\nu] &= 0, \\
    \hnabla_\mu p_\alpha = \hnabla_\mu g_{\alpha \beta} = \vnabla{}^\mu g_{\alpha \beta} &= 0.
\end{align}
\end{subequations}

\section{Eigenspinors -- massive case} \label{app:massive_eigenspinors}

For the case of massive Dirac fields, one generally uses the Dirac basis for spinors and gamma matrices. In this case, the flat spacetime gamma matrices can be written as
\begin{equation}
    \gamma^0 = \begin{pmatrix} -\mathbb{I}_2 & 0 \\ 0 & \mathbb{I}_2 \end{pmatrix}, \qquad \gamma^i = \begin{pmatrix} 0 & \sigma^i \\ -\sigma^i & 0 \end{pmatrix},
\end{equation}
where $\sigma^i$, with $i$ running from $1$ to $3$, are the Pauli matrices defined as
\begin{equation}
    \sigma^1 = \begin{pmatrix} 0 & 1 \\ 1 & 0 \end{pmatrix}, \ \ \sigma^2 = \begin{pmatrix} 0 & -i \\ i & 0 \end{pmatrix}, \ \ \sigma^3 = \begin{pmatrix} 1 & 0 \\ 0 & -1 \end{pmatrix}.
\end{equation}
Working in the Dirac basis, one can construct eigenspinors $\{ \Sigma_0, \Sigma_1, \Pi_0, \Pi_1 \}$ of the principal symbol matrix
\begin{equation}
    D = -\gamma^\mu k_\mu - m \mathbb{I}_4,    
\end{equation}
such that 
\begin{subequations}
\begin{alignat}{3}
    D \Sigma_A &= 0,   &&\bar{\Sigma}_A D = 0, \\
    D \Pi_A &= -2m \Pi_A, \qquad  &&\bar{\Pi}_A D =-2m \bar{\Pi}_A.
\end{alignat}
\end{subequations}
The above equations can be used to show that the following useful relations hold:
\begin{subequations}
\begin{align}
    \Bar{\Sigma}_A \gamma^\mu \Sigma_B &= \frac{k^\mu}{m} \delta_{A B},\\
    \Bar{\Pi}_A \gamma^\mu \Pi_B &= \frac{k^\mu}{m} \delta_{A B}, \\
    \Bar{\Sigma}_A \gamma^\mu \Pi_B &= -i \frac{k_\nu}{m} \Bar{\Sigma}_A \sigma^{\mu \nu} \Pi_B, \\
    \Bar{\Pi}_A \gamma^\mu \Sigma_B &= i \frac{k_\nu}{m} \Bar{\Pi}_A \sigma^{\mu \nu} \Sigma_B.
\end{align}
\end{subequations}
As a concrete example, we present here one possible choice of eigenspinors:
\begin{subequations}
\begin{align}
    \Sigma_0 &= \sqrt{\frac{m - k_0}{2m}}\begin{pmatrix} 1 \\ 0 \\ \frac{k_3}{m-k_0} \\ \frac{k_1 + i k_2}{m - k_0}  \end{pmatrix}, \\
    \Sigma_1 &= \sqrt{\frac{m - k_0}{2m}}\begin{pmatrix} 0 \\ 1 \\ \frac{k_1 - i k_2}{m - k_0} \\ \frac{-k_3}{m-k_0}  \end{pmatrix}, \\
    \Pi_0 &= \sqrt{\frac{m - k_0}{2m}}\begin{pmatrix} \frac{k_3}{m-k_0} \\ \frac{k_1 + i k_2}{m - k_0} \\ 1 \\ 0  \end{pmatrix}, \\
    \Pi_1 &= \sqrt{\frac{m - k_0}{2m}}\begin{pmatrix} \frac{k_1 - i k_2}{m - k_0} \\ \frac{-k_3}{m-k_0} \\ 0 \\ 1  \end{pmatrix}.
\end{align}
\end{subequations}
In the above equations, all vector components are tetrad components $k_a = (e_a)^\mu k_\mu$. Furthermore, the eigenspinors were derived under the restriction that the dispersion relation $k_\mu k^\mu = - m^2$ is satisfied and $k^\mu$ is future-directed with respect to $(e_0)^\mu$. This means that we have $k_0 = - \sqrt{m^2 + (k_1)^2 + (k_2)^2 + (k_3)^2}$. It can easily be checked that these eigenspinors satisfy the orthogonality relations $\bar{\Sigma}_A \Sigma_B = -\bar{\Pi}_A \Pi_B = \delta_{A B}$, $\bar{\Sigma}_A \Pi_B = 0$, and the completeness relation $\Sigma_A \bar{\Sigma}_A - \Pi_A \bar{\Pi}_A = \mathbb{I}_4$.

Note that, except for equations involving the Berry connection, the results presented in Sec. \ref{sec:massive} do not depend on a particular choice of eigenspinors.

\section{Berry curvature -- massive case} \label{app:curvature}

We use the definition of the covariant derivative for spinor fields given in Eq. \eqref{eq:cov_der}, and we can rewrite the components of the Berry curvature as
\begin{subequations}
\begin{align}
    \left( \mathcal{F}_{x x} \right)_{\mu \nu} &= \nabla_\mu \left( \mathcal{B}_x \right)_\nu - \nabla_\nu \left( \mathcal{B}_x \right)_\mu - i \left[ (\mathcal{B}_x)_\mu, (\mathcal{B}_x)_\nu \right], \\
    \left( \mathcal{F}_{k k} \right)^{\mu \nu} &= \vnabla{}^\mu \left( \mathcal{B}_k \right)^\nu  - \vnabla{}^\nu \left( \mathcal{B}_k \right)^\mu - i \left[ (\mathcal{B}_k)^\mu, (\mathcal{B}_k)^\nu \right], \\
    \left( \mathcal{F}_{k x} \right)\indices{_\mu^\nu} &= \nabla_\mu \left( \mathcal{B}_k \right)^\nu - \vnabla{}^\nu \left( \mathcal{B}_x \right)_\mu - i \left[ (\mathcal{B}_x)_\mu, (\mathcal{B}_k)^\nu \right].
\end{align}
\end{subequations}
We insert into the above equations the definition of the components of the Berry connection
\begin{subequations}
\begin{align}
    \left( \mathcal{B}_{x} \right)_\mu &= \frac{i}{2} \left( \bar{\Sigma}_A \nabla_\mu \Sigma_B - \nabla_\mu \bar{\Sigma}_A \Sigma_B \right), \\
    \left( \mathcal{B}_{k} \right)^\mu &= \frac{i}{2} \left( \bar{\Sigma}_A \vnabla{}^\mu \Sigma_B - \vnabla{}^\mu \bar{\Sigma}_A \Sigma_B \right),
\end{align}
\end{subequations}
and we use the orthogonality properties of the eigenspinors:
\begin{equation} \label{eq:orth_derivative}
    \bar{\Sigma}_A \Sigma_B = \delta_{A B} \Rightarrow \left\{
    \begin{array}{ll}
        \nabla_\mu \bar{\Sigma}_A \Sigma_B + \bar{\Sigma}_A \nabla_\mu \Sigma_B = 0, \\
        \vnabla{}^\mu \bar{\Sigma}_A \Sigma_B + \bar{\Sigma}_A \vnabla{}^\mu \Sigma_B = 0.
    \end{array}
\right.
\end{equation}
We obtain
\begin{subequations}
\begin{align}
    \begin{split}
    \left( \mathcal{F}_{x x} \right)_{\mu \nu} &= i(\bar{\Sigma}_A \nabla_{[\mu} \nabla_{\nu]} \Sigma_B - \nabla_{[\mu} \nabla_{\nu]} \bar{\Sigma}_A \Sigma_B ) \\
    & \quad+ 2i (\nabla_{[\mu} \bar{\Sigma}_A) (\mathbb{I}_4 - \Sigma_C \bar{\Sigma}_C) (\nabla_{\nu]} \Sigma_B), 
    \end{split} \\
    \begin{split}
    \left( \mathcal{F}_{k k} \right)^{\mu \nu} &= 2i (\vnabla{}^{[\mu} \bar{\Sigma}_A) (\mathbb{I}_4 - \Sigma_C \bar{\Sigma}_C) (\vnabla{}^{\nu]} \Sigma_B), 
    \end{split} \\
    \begin{split}
    \left( \mathcal{F}_{k x} \right)\indices{_\mu^\nu} &= 2i (\nabla_{[\mu} \bar{\Sigma}_A) (\mathbb{I}_4 - \Sigma_C \bar{\Sigma}_C) (\vnabla{}^{\nu]} \Sigma_B). 
    \end{split}
\end{align}
\end{subequations}
Furthermore, we use the resolution of identity given in Eq. \eqref{eq:Res_id}, and we obtain
\begin{subequations}
\begin{align}
    \left( \mathcal{F}_{x x} \right)_{\mu \nu} &= i(\bar{\Sigma}_A \nabla_{[\mu} \nabla_{\nu]} \Sigma_B - \nabla_{[\mu} \nabla_{\nu]} \bar{\Sigma}_A \Sigma_B ) \\
    &\quad - 2i (\nabla_{[\mu} \bar{\Sigma}_A) \Pi_C \bar{\Pi}_C (\nabla_{\nu]} \Sigma_B), \\
    \left( \mathcal{F}_{k k} \right)^{\mu \nu} &= -2i (\vnabla{}^{[\mu} \bar{\Sigma}_A) \Pi_C \bar{\Pi}_C (\vnabla{}^{\nu]} \Sigma_B), \\
    \left( \mathcal{F}_{k x} \right)\indices{_\mu^\nu} &= -2i (\nabla_{[\mu} \bar{\Sigma}_A) \Pi_C \bar{\Pi}_C (\vnabla{}^{\nu]} \Sigma_B).
\end{align}
\end{subequations}
The commutator of spinor covariant derivatives can be expressed in terms of the Riemann tensor as
\begin{equation}
\begin{split}
    (\nabla_\mu \nabla_\nu - \nabla_\nu \nabla_\mu) \Psi &= -\frac{1}{4} R_{\mu \nu \rho \sigma} \gamma^\rho \gamma^\sigma \Psi, \\ 
    (\nabla_\mu \nabla_\nu - \nabla_\nu \nabla_\mu) \bar{\Psi} &= \frac{1}{4} R_{\mu \nu \rho \sigma} \bar{\Psi} \gamma^\rho \gamma^\sigma.
\end{split}
\end{equation}
Using these relations, we can write
\begin{equation}
    i(\bar{\Sigma}_A \nabla_{[\mu} \nabla_{\nu]} \Sigma_B - \nabla_{[\mu} \nabla_{\nu]} \bar{\Sigma}_A \Sigma_B ) = -\frac{1}{2} R_{\mu \nu \alpha \beta} s^{\alpha \beta}.
\end{equation}
The remaining terms that need to be computed are of the form $\bar{\Pi}_C \nabla_\mu \Sigma_B$ or $\bar{\Pi}_C \vnabla{}^\mu \Sigma_B$. For this purpose, we can use the fact that $\Sigma_A$ is an eigenspinor of $D$, with eigenvalue zero, while $\Pi_A$ is an eigenspinor of $D$, with eigenvalue $-2m$. We start with
\begin{equation}
    \bar{\Pi}_A D \Sigma_B = 0.
\end{equation}
Taking a vertical derivative of this expression, we obtain
\begin{equation}
\begin{split}
    0 &= (\vnabla{}^\mu \bar{\Pi}_A) D \Sigma_B + \bar{\Pi}_A (\vnabla{}^\mu D) \Sigma_B + \bar{\Pi}_A D (\vnabla{}^\mu \Sigma_B) \\
    &= \bar{\Pi}_A (\vnabla{}^\mu D) \Sigma_B -2 m \bar{\Pi}_A (\vnabla{}^\mu \Sigma_B) \\
    &= -\bar{\Pi}_A \gamma^\mu \Sigma_B -2 m \bar{\Pi}_A (\vnabla{}^\mu \Sigma_B),
\end{split}
\end{equation}
and we can finally write
\begin{equation} \label{eq:term_1}
    \bar{\Pi}_A \vnabla{}^\mu \Sigma_B = -\frac{1}{2m} \bar{\Pi}_A \gamma^\mu \Sigma_B
\end{equation}
Similarly, taking a covariant derivative of $\bar{\Pi}_A D \Sigma_B = 0$, we obtain
\begin{equation}
\begin{split}
    0 &= (\nabla_\mu \bar{\Pi}_A) D \Sigma_B + \bar{\Pi}_A (\nabla_\mu D) \Sigma_B + \bar{\Pi}_A D (\nabla_\mu \Sigma_B) \\
    &= \bar{\Pi}_A (\nabla_\mu D) \Sigma_B -2 m \bar{\Pi}_A (\nabla_\mu \Sigma_B) \\
    &= -(\nabla_\mu k_\alpha) \bar{\Pi}_A \gamma^\alpha \Sigma_B -2 m \bar{\Pi}_A (\nabla_\mu \Sigma_B).
\end{split}
\end{equation}
However, since we are working on $T^*M$, with coordinates $(x^\mu, k_\mu)$, we have
\begin{equation}
    \nabla_\mu k_\alpha = \frac{\partial}{\partial x^\mu} k_\alpha - \Gamma^\sigma_{\mu \alpha} k_\sigma = - \Gamma^\sigma_{\mu \alpha} k_\sigma.
\end{equation}
Thus, we can write
\begin{equation} \label{eq:term_2}
\begin{split}
    \bar{\Pi}_A \nabla_\mu \Sigma_B &= -\frac{1}{2m} (\nabla_\mu k_\alpha) \bar{\Pi}_A \gamma^\alpha \Sigma_B \\
    &= \frac{1}{2m} k_\sigma \Gamma^\sigma_{\mu \alpha} \bar{\Pi}_A \gamma^\alpha \Sigma_B.
\end{split}
\end{equation}
Using Eqs. \eqref{eq:term_1} and \eqref{eq:term_2}, we arrive at the final form for the components of the Berry curvature:
\begin{align}
    \left( \mathcal{F}_{x x} \right)_{\mu \nu} &= -\frac{1}{2} R_{\mu \nu \alpha \beta} s^{\alpha \beta} + \frac{1}{m^2} k_\rho k_\sigma \Gamma^\rho_{\alpha \mu} \Gamma^\sigma_{\beta \nu} s^{\alpha \beta}, \\
    \left( \mathcal{F}_{k k} \right)^{\mu \nu} &= \frac{1}{m^2} s^{\mu \nu}, \\
    \left( \mathcal{F}_{k x} \right)\indices{_\mu^\nu} &= -\left( \mathcal{F}_{x k} \right)\indices{^\nu_\mu} =- \frac{1}{m^2} k_\rho \Gamma^\rho_{\mu \alpha} s^{\alpha \nu}.
\end{align}

\begin{widetext}
\section{Other calculations -- massive case} \label{app:noname}

Using the definitions of $s^{\alpha \beta}$ and $(\mathcal{B}_k)^\mu$, we can rewrite the last term in Eq. \eqref{eq:dotX_2} as
\begin{equation}
\begin{split}
     \vnabla{}^\mu s^{\alpha \beta} - i [(\mathcal{B}_k)^\mu, s^{\alpha \beta}] &= \frac{1}{2} \left[ (\vnabla{}^\mu \bar{\Sigma}_A) \sigma^{\alpha \beta} \Sigma_B + \bar{\Sigma}_A \sigma^{\alpha \beta} (\vnabla{}^\mu \Sigma_B) \right] + \frac{1}{4} \left[ \bar{\Sigma}_A (\vnabla{}^\mu \Sigma_C) - (\vnabla{}^\mu \bar{\Sigma}_A) \Sigma_C \right] \bar{\Sigma}_C \sigma^{\alpha \beta} \Sigma_B \\
     &\qquad - \frac{1}{4} \bar{\Sigma}_A \sigma^{\alpha \beta} \Sigma_C \left[ \bar{\Sigma}_C (\vnabla{}^\mu \Sigma_B) - (\vnabla{}^\mu \bar{\Sigma}_C) \Sigma_B \right] \\
     &= \frac{1}{4} \left[ 2(\vnabla{}^\mu \bar{\Sigma}_A) + \bar{\Sigma}_A (\vnabla{}^\mu \Sigma_C) \bar{\Sigma}_C - (\vnabla{}^\mu \bar{\Sigma}_A) \Sigma_C \bar{\Sigma}_C \right] \sigma^{\alpha \beta} \Sigma_B \\
     &\qquad + \frac{1}{4} \bar{\Sigma}_A \sigma^{\alpha \beta} \left[ 2(\vnabla{}^\mu \Sigma_B) - \Sigma_C \bar{\Sigma}_C (\vnabla{}^\mu \Sigma_B) + \Sigma_C (\vnabla{}^\mu \bar{\Sigma}_C) \Sigma_B \right].
\end{split}
\end{equation}
Using Eqs. \eqref{eq:orth_derivative} and \eqref{eq:Res_id}, the above expression simplifies to
\begin{equation}
\begin{split}
     \vnabla{}^\mu s^{\alpha \beta} - i [(\mathcal{B}_k)^\mu, s^{\alpha \beta}] &= \frac{1}{2} \left[ (\vnabla{}^\mu \bar{\Sigma}_A) - (\vnabla{}^\mu \bar{\Sigma}_A) \Sigma_C \bar{\Sigma}_C \right] \sigma^{\alpha \beta} \Sigma_B + \frac{1}{2} \bar{\Sigma}_A \sigma^{\alpha \beta} \left[ (\vnabla{}^\mu \Sigma_B) - \Sigma_C \bar{\Sigma}_C (\vnabla{}^\mu \Sigma_B) \right] \\
     &= \frac{1}{2} (\vnabla{}^\mu \bar{\Sigma}_A) ( \mathbb{I}_4 - \Sigma_C \bar{\Sigma}_C ) \sigma^{\alpha \beta} \Sigma_B + \frac{1}{2} \bar{\Sigma}_A \sigma^{\alpha \beta} ( \mathbb{I}_4 - \Sigma_C \bar{\Sigma}_C ) (\vnabla{}^\mu \Sigma_B) \\
      &= -\frac{1}{2} (\vnabla{}^\mu \bar{\Sigma}_A) \Pi_C \bar{\Pi}_C \sigma^{\alpha \beta} \Sigma_B - \frac{1}{2} \bar{\Sigma}_A \sigma^{\alpha \beta} \Pi_C \bar{\Pi}_C (\vnabla{}^\mu \Sigma_B).
\end{split}
\end{equation}
Inserting Eq. \eqref{eq:term_1} and its complex conjugate into the above expression, we obtain
\begin{equation}
\begin{split}
     \vnabla{}^\mu s^{\alpha \beta} - i [(\mathcal{B}_k)^\mu, s^{\alpha \beta}] &= \frac{1}{4m} \bar{\Sigma}_A \gamma^\mu \Pi_C \bar{\Pi}_C \sigma^{\alpha \beta} \Sigma_B + \frac{1}{4m} \bar{\Sigma}_A \sigma^{\alpha \beta} \Pi_C \bar{\Pi}_C \gamma^\mu \Sigma_B \\
     &= \frac{1}{4m} \bar{\Sigma}_A \gamma^\mu (\Sigma_C \bar{\Sigma}_C - \mathbb{I}_4) \sigma^{\alpha \beta} \Sigma_B + \frac{1}{4m} \bar{\Sigma}_A \sigma^{\alpha \beta} (\Sigma_C \bar{\Sigma}_C - \mathbb{I}_4) \gamma^\mu \Sigma_B \\
     &= \frac{1}{4m} \bar{\Sigma}_A \gamma^\mu (\Sigma_C \bar{\Sigma}_C - \mathbb{I}_4) \sigma^{\alpha \beta} \Sigma_B + \frac{1}{4m} \bar{\Sigma}_A \sigma^{\alpha \beta} (\Sigma_C \bar{\Sigma}_C - \mathbb{I}_4) \gamma^\mu \Sigma_B \\
     &= \frac{1}{ m^2} P^\mu s^{\alpha \beta} - \frac{1}{4m} \bar{\Sigma}_A (\gamma^\mu \sigma^{\alpha \beta} + \sigma^{\alpha \beta} \gamma^\mu) \Sigma_B .
\end{split}
\end{equation}
The anticommutator $\gamma^\mu \sigma^{\alpha \beta} + \sigma^{\alpha \beta} \gamma^\mu$ can be rewritten in a different form by using the properties of the gamma matrices. We consider the flat spacetime gamma matrices $\gamma^a$, which are related to the spacetime gamma matrices by the orthonormal tetrad as $\gamma^\mu = (e_a)^\mu \gamma^a$. For the flat spacetime gamma matrices, we can write the following relation:
\begin{equation}
    \gamma^a \gamma^b \gamma^c = - \eta^{a b} \gamma^c - \eta^{b c} \gamma^a + \eta^{a c} \gamma^b + i \epsilon^{d a b c} \gamma_d \gamma^5,
\end{equation}
where $\eta^{a b}$ is the Minkowski metric tensor, with signature $(-\,+\,+\,+)$, and $\gamma^5 = i \gamma^0 \gamma^1 \gamma^2 \gamma^3$. Using this relation, we obtain (this is the spin density tensor; see Refs. \citep{Hehl1971,Turcati2020})
\begin{equation}
    \gamma^c \sigma^{a b} + \sigma^{a b} \gamma^c = 2 \epsilon^{d a b c} \gamma_d \gamma^5
\end{equation}
and
\begin{equation}
    \gamma^\mu \sigma^{\alpha \beta} + \sigma^{\alpha \beta} \gamma^\mu = 2 (e_a)^\alpha (e_b)^\beta (e_c)^\mu \epsilon^{d a b c} \gamma_d \gamma^5 = 2 (e_a)^\alpha (e_b)^\beta (e_c)^\mu (e_d)^\nu \epsilon^{d a b c} \gamma_\nu \gamma^5.
\end{equation}
Furthermore, it can be shown that \citep[Eq. 4.12]{audretsch_torsion}
\begin{equation}
    \bar{\Sigma}_A \gamma_a \gamma^5 \Sigma_B = -\frac{1}{2m} \epsilon_{a b c d} \bar{\Sigma}_A \sigma^{b c} \Sigma_B P^d = -\frac{1}{m} \epsilon_{a b c d} s^{b c} P^d.
\end{equation}
This relation can be inverted by using the properties of the Levi-Civita tensor:
\begin{equation}
    \epsilon^{d a b c} \bar{\Sigma}_A \gamma_d \gamma^5 \Sigma_B = \frac{2}{m} \left( s^{a b} P^c + s^{c a} P^b + s^{b c} P^a \right).
\end{equation}
Thus, we can finally write
\begin{equation}
\begin{split}
     \vnabla{}^\mu s^{\alpha \beta} - i [(\mathcal{B}_k)^\mu, s^{\alpha \beta}] &= - \frac{1}{m^2} \left( s^{\mu \alpha} P^\beta + s^{\beta \mu} P^\alpha \right).
\end{split}
\end{equation}

We can perform a similar calculation for the last term in Eq. \eqref{eq:dotP_2}. First, note that by using the properties of the covariant derivative, we can write
\begin{equation}
    F_{\alpha \beta, \mu} \bar{z} s^{\alpha \beta}z + F_{\alpha \beta} \bar{z}  s\indices{^{\alpha \beta}_{,\mu}} z = (\nabla_\mu F_{\alpha \beta}) \bar{z} s^{\alpha \beta}z + F_{\alpha \beta} \bar{z} \nabla_\mu s^{\alpha \beta} z.
\end{equation}
The last term in Eq. \eqref{eq:dotP_2} becomes $\nabla_\mu s^{\alpha \beta} - i [(\mathcal{B}_x)_\mu, s^{\alpha \beta}]$, and we can apply the same steps as before. We obtain
\begin{equation}
    \nabla_\mu s^{\alpha \beta} - i [(\mathcal{B}_x)_\mu, s^{\alpha \beta}] = -\frac{1}{2} (\nabla_\mu \bar{\Sigma}_A) \Pi_C \bar{\Pi}_C \sigma^{\alpha \beta} \Sigma_B - \frac{1}{2} \bar{\Sigma}_A \sigma^{\alpha \beta} \Pi_C \bar{\Pi}_C (\nabla_\mu \Sigma_B).
\end{equation}
Using Eq. \eqref{eq:term_2} and its complex conjugate, we obtain
\begin{equation}
\begin{split}
    \nabla_\mu s^{\alpha \beta} - i [(\mathcal{B}_x)_\mu, s^{\alpha \beta}] &= -\frac{1}{4m} P_\sigma \Gamma^\sigma_{\mu \rho} \left( \bar{\Sigma}_A \gamma^\rho \Pi_C \bar{\Pi}_C \sigma^{\alpha \beta} \Sigma_B + \bar{\Sigma}_A \sigma^{\alpha \beta} \Pi_C \bar{\Pi}_C \gamma^\rho \Sigma_B \right) \\
    &= -\frac{1}{m^2}  P_\sigma P^\rho \Gamma^\sigma_{\mu \rho} s^{\alpha \beta} + \frac{1}{4m}  P_\sigma \Gamma^\sigma_{\mu \rho} \bar{\Sigma}_A (\gamma^\rho \sigma^{\alpha\beta} +\sigma^{\alpha\beta} \gamma^\rho) \Sigma_B.
\end{split}
\end{equation}

\end{widetext}

\section{Eigenspinors -- massless case} \label{app:massless_eigenspinors}

In this section, we discuss the properties of the eigenspinors of the principal symbol matrix $D_0$, and we derive some useful relations. As an example, we also give a particular choice of eigenspinors.

For massless Dirac fields, it is more convenient to work in the Weyl (or chiral) basis. One advantage is that the transition between $4$-spinors and $2$-spinors is more transparent. The gamma matrices are defined as
\begin{equation}
    \gamma^a = \begin{pmatrix} 0 & \sigma^a \\ \bar{\sigma}^a & 0 \end{pmatrix},
\end{equation}
where $\sigma^a = (\mathbb{I}_2, \sigma^i)$, $\bar{\sigma}^a = (\mathbb{I}_2, -\sigma^i)$, and $\sigma^i$ are the Pauli matrices. Note that the bar in $\bar{\sigma}^a$ is used only for notation and does not represent complex conjugation.

Under the restrictions of the dispersion relation $k_\mu k^\mu = 0$ and $k^\mu$ future-directed, meaning that $k_0 = - \sqrt{(k_1)^2 + (k_2)^2 + (k_3)^2}$, the principal symbol matrix
\begin{equation}
    D_0 = -\gamma^\mu k_\mu    
\end{equation}
admits two eigenspinors of eigenvalue zero, which we can write as
\begin{equation}
    \Sigma_0 = \begin{pmatrix} 0 \\ u \end{pmatrix}, \qquad  \Sigma_1 = \begin{pmatrix} v \\ 0 \end{pmatrix}.
\end{equation}
The $2$-spinors $u$ and $v$ must satisfy the following relations:
\begin{subequations} \label{eq:eigen2spinors}
\begin{align}
    k_\mu \sigma^\mu u = 0, \qquad k_\mu \sigma^\mu v = 2k_0 v, \\
    k_\mu \bar{\sigma}^\mu v = 0, \qquad k_\mu \bar{\sigma}^\mu u = 2k_0 u.
\end{align}
\end{subequations}
The $2$-spinors also satisfy the orthogonality relations
\begin{equation} \label{eq:2spinor_ortho}
    \bar{u}.u = \bar{v}.v = 1, \qquad \bar{u}.v = \bar{v}.u = 0,
\end{equation}
and the completeness relation
\begin{equation}
    u \bar{u} + v \bar{v} = \mathbb{I}_2.
\end{equation}
The following relations will also be used:
\begin{subequations}
\begin{align}
    k_a \sigma^a u &= 0 \qquad \Rightarrow \qquad u = - \frac{k_i \sigma^i u}{k_0}, \\
    k_a \bar{u} \sigma^a &= 0 \qquad \Rightarrow \qquad \bar{u} = - \frac{k_i \bar{u} \sigma^i}{k_0}, \\
    k_a \bar{\sigma}^a v &= 0 \qquad \Rightarrow \qquad v = \frac{k_i \sigma^i v}{k_0}, \\
    k_a \bar{v} \bar{\sigma}^a &= 0 \qquad \Rightarrow \qquad \bar{v} = \frac{k_i \bar{v} \sigma^i}{k_0}.
\end{align}
\end{subequations}
Using the above equations, one can show that 
\begin{equation} \label{eq:2spinor_momentum}
    \bar{u} \sigma^\mu u = \bar{v} \bar{\sigma}^\mu v = \frac{k^\mu}{t_\alpha k^\alpha}.
\end{equation}
In general, two eigenspinors $u$ and $v$ satisfying the above relations can always be related as
\begin{equation} \label{eq:conj}
    u = e^{i \theta} \sigma_2 v^*,
\end{equation}
for some choice of $\theta \in \mathbb{R}$. This relation is related to charge conjugation.

As an example, here we give a particular choice of $u$ and $v$. One possibility is 
\begin{subequations}
\begin{align}
    u &= \frac{1}{\sqrt{2\left[(k_0)^2 - k_0 k_3 \right]}} \begin{pmatrix} k_3 - k_0 \\ k_1 + i k_2 \end{pmatrix}, \\
    v &= \frac{1}{\sqrt{2\left[(k_0)^2 + k_0 k_3 \right]}} \begin{pmatrix} k_3 + k_0 \\ k_1 + i k_2 \end{pmatrix}.
\end{align}
\end{subequations}
It can be checked that the above relations are satisfied with this particular choice of eigenspinors.

\newpage

\bibliography{references}

\end{document}